\documentclass[twocolumn,secnumarabic,amssymb, nobibnotes, aps, prd,longbibliography]{revtex4-2}

\usepackage{graphicx}  
\usepackage{dcolumn}   
\usepackage{bm}        
\usepackage{verbatim}   
\usepackage{graphicx}
\usepackage{epstopdf}
\usepackage{footnote}
\usepackage{epsfig}
\usepackage{subfigure}
\usepackage{amssymb}
\usepackage{amsmath,bm}
\usepackage{xcolor}
\usepackage{float}
\usepackage{subfigure}
\usepackage[most]{tcolorbox}
\usepackage{soul}
\usepackage{color}
\usepackage[colorlinks=true,linkcolor=blue,allcolors=blue]{hyperref}%
\usepackage[framemethod=TikZ]{mdframed}
\usepackage{lipsum}
\mdfdefinestyle{MyFrame}{%
	linecolor=blue,
	outerlinewidth=2pt,
	roundcorner=20pt,
	innertopmargin=\baselineskip,
	innerbottommargin=\baselineskip,
	innerrightmargin=20pt,
	innerleftmargin=20pt,
	backgroundcolor=gray!50!white}

\begin{document}
	
	
	\headsep = 40pt
\title{Bianisotropic Space-Time Metasurfaces}
	\author{Sajjad Taravati* and Ya-Adama Kabia}
	
	\affiliation{School of Electronics and Computer Science, University of Southampton, Southampton SO17 1BJ, UK}
	\email{e-mail: s.taravati@soton.ac.uk}
	
\begin{abstract}
We experimentally demonstrate bianisotropic space-time metasurfaces based on varactor-loaded omega-shaped meta-atoms. The meta-atoms support coexisting electric and magnetic responses, with pronounced electric-field distributions and strong ring-associated magnetic fields at the operating frequency. A traveling-wave modulation drives reactive loads embedded in both responsive regions, rendering the composite bianisotropic response periodic in space and time. The resulting modulation generates a reconfigurable polychromatic spectrum with directional sideband radiation governed by Floquet phase matching. Under free-space illumination, the same metasurface preferentially couples the received field toward one guided port, producing modulation-controlled directional routing. These results establish bianisotropic space-time metasurfaces as a platform for jointly controlling frequency generation, directional radiation, and wave routing within a single dynamically modulated aperture.
\end{abstract}
	
\maketitle

\section{Introduction}\label{sec:introduction}
\begin{figure*}
		\begin{center}
	\includegraphics[width=2\columnwidth]{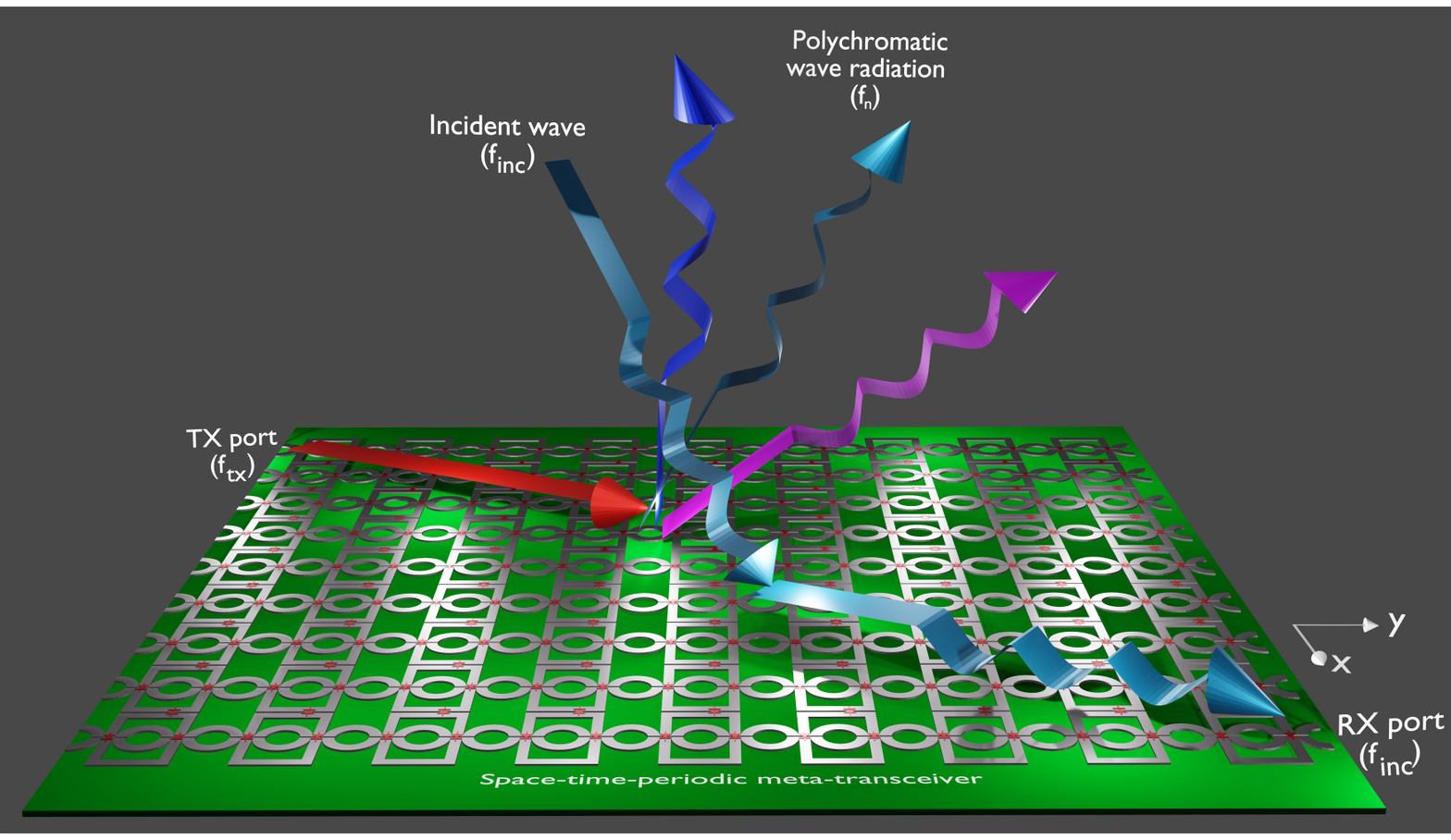}
\caption{Operating principle of the bianisotropic space-time metasurface. A low-frequency excitation at angular frequency $\omega_{\mathrm{tx}}$ interacts with a traveling modulation of angular frequency $\Omega$, generating space-time harmonics $\omega_{n,m}=n\Omega+m\omega_{\mathrm{tx}}$. Radiative harmonics couple to free space according to their Floquet phase-matching conditions. Under free-space illumination, the same traveling modulation produces preferential coupling toward the RX channel, providing directional receive-port routing.}
			\label{Fig:1}
		\end{center}
	\end{figure*}

Dynamic metamaterials and metasurfaces provide a versatile platform for controlling the frequency and momentum of electromagnetic waves~\cite{Taravati_Kishk_MicMag_2019}, enabling functionalities such as directional radiation~\cite{taravati2016mixer,karimian2021nonreciprocal,Taravati_AMA_PRApp_2020}, frequency conversion~\cite{Taravati_PRB_Mixer_2018,taravati2021pure}, parametric amplification~\cite{Cullen_NAT_1958,Taravati_NC_2021}, and direction-dependent coupling between spatial channels~\cite{bharadia2013full,sabharwal2014band,salary2019nonreciprocal,zhu2023joint,taravati2020full}. Among these approaches, space-time modulation is particularly attractive because a traveling modulation simultaneously supplies temporal frequency translation and spatial momentum. It can therefore couple electromagnetic states separated in both frequency and momentum, providing a magnet-free route to frequency conversion, directional radiation, and nonreciprocal wave interactions~\cite{taravati2016mixer,Taravati_AMA_PRApp_2020,karimian2021nonreciprocal,valizadeh2025integrated}. Most experimental space-time metasurfaces reported to date have relied predominantly on electrically responsive elements, such as varactor-loaded dipoles and patches~\cite{taravati2020full,zang2019nonreciprocal,taravati2021pure,10821482}. An important extension is obtained when the meta-atom supports coupled electric and magnetic responses. In a bianisotropic medium, the induced electric and magnetic polarizations are interconnected through magnetoelectric coupling, providing additional degrees of freedom beyond a purely electric space-time modulation. Although time-dependent electric and magnetic constitutive responses have been investigated theoretically~\cite{Taravati_PRAp_2018,Taravati_Kishk_TAP_2019,huidobro2019fresnel,gaxiola2021temporal,taravati2025finite}, their experimental implementation in a dynamically modulated metasurface remains considerably less explored. Here, we experimentally demonstrate a bianisotropic space-time metasurface formed by varactor-loaded omega-shaped meta-atoms. The interconnected dipolar and ring sections support coexisting electric and magnetic responses at the operating frequency. In particular, the electric field exhibits pronounced transverse localization within the circular rings, while strong magnetic fields and magnetic-energy localization occur in the same ring regions. The omega geometry therefore supports a coupled electric--magnetic electromagnetic state rather than a purely electric dipolar response. Varactors embedded in both responsive regions are driven by a common traveling-wave bias, rendering the composite bianisotropic response periodic in space and time. The resulting metasurface couples the excitation to a set of space-time Floquet harmonics, $\omega_{n,m}=n\Omega+m\omega_{\mathrm{tx}}$, whose spectral and spatial properties are jointly determined by the temporal and spatial momenta supplied by the modulation. Experimentally, this interaction produces a bias-reconfigurable polychromatic spectrum and directional radiation of the generated sidebands. The first mixing sidebands occupy approximately the same angular region, consistent with their common spatial Floquet order, while free-space illumination of the same modulated aperture preferentially couples the received field toward one guided port. These observations demonstrate how bianisotropy and traveling space-time modulation can be combined within a single metasurface to control frequency generation, radiation, and spatial routing. Figure~\ref{Fig:1} summarizes the operating principle. A low-frequency excitation at angular frequency $\omega_{\mathrm{tx}}$ interacts with the traveling modulation at angular frequency $\Omega$ to generate the polychromatic spectrum $\omega_{n,m}=n\Omega+m\omega_{\mathrm{tx}}$. Radiating Floquet states couple to free space according to their phase-matching conditions, whereas under free-space illumination the traveling modulation produces preferential coupling toward the RX channel. The following sections develop the bianisotropic constitutive description and Floquet model and then examine the electric and magnetic near fields, fabricated metasurface, generated spectra, directional radiation, and receive-port routing.

\section{Theoretical Framework}\label{sec:theory}
Throughout this paper, angular-frequency symbols are used in the formulation to avoid ambiguity between the modulation label and the integer mixing index $m$. The traveling-wave modulation has angular frequency $\Omega$, with $\Omega$, and the TX excitation has angular frequency $\omega_\text{tx}$, with $\omega_\text{tx}/(2\pi)=10$~MHz. Spectral components are indexed as $\omega_{n,m}=n\Omega+m\omega_\text{tx}$; ordinary frequencies in gigahertz are reported only as $\omega_{n,m}/(2\pi)$.

Figure~\ref{Fig:1} summarizes the two measured operating configurations. In transmit operation, the TX-port excitation $\omega_\text{tx}$ and the traveling-wave modulation $\Omega$ generate a set of mixing products around the pump frequency, some of which radiate into free space. In receive operation, free-space illumination at $\omega_\text{inc}$ produces different powers at the RX and TX ports under the same bias and incidence conditions. The latter observation is reported as directional port routing. Although traveling space-time modulation breaks time-reversal symmetry in the underlying medium, the present port comparison is not treated as a complete experimental reciprocity test.

\subsection{Bianisotropic Unit-Cell Response}\label{subsec:unitcell_response}
For a tangentially polarizable bianisotropic sheet, the electric and magnetic surface polarizations may be written in compact form as
\begin{equation}
\begin{bmatrix}
\mathbf{P}_\text{t}\\[2pt]
\mathbf{M}_\text{t}
\end{bmatrix}
=
\begin{bmatrix}
\epsilon_0\overline{\overline{\chi}}_\text{ee} & c^{-1}\overline{\overline{\chi}}_\text{em}\\[2pt]
\eta_0^{-1}\overline{\overline{\chi}}_\text{me} & \overline{\overline{\chi}}_\text{mm}
\end{bmatrix}
\begin{bmatrix}
\mathbf{E}_\text{av}\\[2pt]
\mathbf{H}_\text{av}
\end{bmatrix}.
\label{eq:bianisotropic_constitutive}
\end{equation}
The omega-shaped geometry couples the electric and magnetic responses through the off-diagonal magnetoelectric terms. Since these tensors are not retrieved in the present work, Eq.~\eqref{eq:bianisotropic_constitutive} provides the constitutive interpretation rather than a fitted model.

The traveling-wave voltage applied to all three varactors is
\begin{equation}
V_j(x,t)=V_\text{DC}+V_\text{mod}\cos\!\left(\Omega t-k_\text{m}x\right),\qquad j\in\{1,2,3\}.
\label{eq:common_bias}
\end{equation}
The corresponding capacitances $C_{D_j}[V_j(x,t)]$ vary periodically in space and time. Because $\text{D}_1$ loads the predominantly electric path and $\text{D}_{2,3}$ load ring paths that carry a pronounced magnetic response, the common bias produces a space-time-varying composite bianisotropic response. In general, each effective susceptibility in Eq.~\eqref{eq:bianisotropic_constitutive} may contain a first-order modulation term,
\begin{equation}
\overline{\overline{\chi}}_{ab}(x,t)=
\overline{\overline{\chi}}_{ab}^{(0)}+
\Delta\overline{\overline{\chi}}_{ab}
\cos\!\left(\Omega t-k_\text{m}x+\phi_{ab}\right),
\quad a,b\in\{e,m\}.
\label{eq:bianisotropic_modulation}
\end{equation}
The amplitudes $\Delta\overline{\overline{\chi}}_{ab}$ and phases $\phi_{ab}$ are not independently extracted. Therefore, the measurements do not establish separate numerical modulation depths for the electric, magnetic, and magnetoelectric components.

For the Floquet and phase-matching analysis, we use a reduced scalar transmission-line model that retains effective electric and magnetic responses while absorbing magnetoelectric coupling into the fitted modal parameters. The reduced susceptibilities are written as
\begin{subequations}
\begin{align}
\chi_\text{ee}(x,t)&=\chi_\text{ee,0}+\chi_\text{ee,1}\cos\!\left(\Omega t-k_\text{m}x\right),\label{eq:chi_ee}\\[2pt]
\chi_\text{mm}(x,t)&=\chi_\text{mm,0}+\chi_\text{mm,1}\cos\!\left(\Omega t-k_\text{m}x\right).\label{eq:chi_mm}
\end{align}
\end{subequations}
Let $V(x,t)\equiv E_\parallel(x,t)$ and $I(x,t)\equiv H_\parallel(x,t)$ denote modal tangential fields. The reduced coupled equations are
\begin{subequations}
\begin{align}
\frac{\partial I}{\partial x}&=-\epsilon_0\frac{\partial}{\partial t}\!\left[\chi_\text{ee}(x,t)V(x,t)\right],\label{eq:gstc_I}\\[4pt]
\frac{\partial V}{\partial x}&=-\mu_0\frac{\partial}{\partial t}\!\left[\chi_\text{mm}(x,t)I(x,t)\right].\label{eq:gstc_V}
\end{align}
\end{subequations}
This model identifies the dipole-loaded path with an effective time-varying shunt response and the ring-loaded path with an effective time-varying series response. It is used to organize the harmonic and phase-matching physics, not to retrieve the full tensorial bianisotropic sheet parameters.

\subsection{Unmodulated Dispersion and Floquet Space-Time Harmonics}\label{subsec:dispersion}
	
	With the modulation switched off ($\chi_\text{ee,1}=\chi_\text{mm,1}=0$),
	Eqs.~\eqref{eq:gstc_I}--\eqref{eq:gstc_V} support a traveling wave
	$V,I \propto \exp\!\big[i(\beta_0 x - \omega t)\big]$ with dispersion
	relation
	\begin{equation}
		\beta_0(\omega) = \omega \sqrt{\mu_0 \epsilon_0\, \chi_\text{mm,0}\,\chi_{ee,0}}.
		\label{eq:beta0}
	\end{equation}
	
	With the modulation active, $\chi_\text{ee}(x,t)$ and $\chi_\text{mm}(x,t)$ are
	periodic in both $x$ and $t$ with the same period as the bias
	waveform, so the Floquet--Bloch theorem applies and the fields expand
	as an infinite sum of space-time harmonics
\begin{subequations}
	\begin{align}
		V(x,t) &= \sum_{n,m=-\infty}^{\infty} V_{n,m}\, e^{\,i(\beta_{n,m} x - \omega_{n,m} t)},
		\label{eq:V_expand} \\[2pt]
		I(x,t) &= \sum_{n,m=-\infty}^{\infty} I_{n,m}\, e^{\,i(\beta_{n,m} x - \omega_{n,m} t)},
		\label{eq:I_expand}
	\end{align}
	with harmonic frequencies and wavenumbers
	\begin{equation}
		\omega_{n,m} = n\Omega + m\,\omega_\text{tx}, \qquad
		\beta_{n,m} = n k_\text{m} + m\,\beta_\text{tx}, \qquad n,m \in \mathbb{Z},
		\label{eq:harmonic_ladder}
	\end{equation}
\end{subequations}
	where $\Omega$, $\omega_\text{tx}$, and $\omega_\text{inc}$ are the angular modulation, TX-excitation, and incident frequencies, respectively. Here
	$\beta_\text{tx}=\beta_0(\omega_\text{tx})$ its guided wavenumber from Eq.~\eqref{eq:beta0}. The index
	$n$ counts quanta of the modulation $(\Omega,k_\text{m})$, and the index $m$ counts harmonics of the
	baseband tone, with $m=\pm1$ its linear (up- and down-converted) response and $|m|\ge2$ its weak
	nonlinear products in the varactors. The measured comb lies in the family $n=1$, at
	$\omega_{1,m}=\Omega+m\omega_\text{tx}$.
	For a single-tone excitation at $\omega_\text{inc}$ and no baseband tone, as in reception, $m$ is dropped and the
	ladder reduces to $\omega_\text{inc}+n\Omega$.
	
	Substituting Eqs.~\eqref{eq:V_expand}--\eqref{eq:I_expand} into
	Eqs.~\eqref{eq:gstc_I}--\eqref{eq:gstc_V}, and using the cosine-form
	Fourier coefficients $\chi_\text{ee}^{(0)} = \chi_\text{ee,0}$,
	$\chi_\text{ee}^{(\pm 1)} = \chi_\text{ee,1}/2$ (and identically for
	$\chi_\text{mm}^{(q)}$), yields the coupled-mode equations governing the
	harmonic amplitudes along the line, for each baseband harmonic $m$
\begin{subequations}
	\begin{align}
		\frac{d I_{n,m}}{dx} &= i\,\epsilon_0 \sum_{q=-1}^{1}
		\chi_\text{ee}^{(q)}\, \omega_{n-q,m}\, V_{n-q,m},
		\label{eq:coupled_I} \\[4pt]
		\frac{d V_{n,m}}{dx} &= i\,\mu_0 \sum_{q=-1}^{1}
		\chi_\text{mm}^{(q)}\, \omega_{n-q,m}\, I_{n-q,m}.
		\label{eq:coupled_V}
	\end{align}
\end{subequations}	
	Truncating to $|n| \le N$ (and $|m|\le M$ when the baseband nonlinearity is retained) gives a finite linear system that can be
	solved numerically for the harmonic amplitudes once
	$\chi_\text{ee,1}(V_\text{DC})$ and $\chi_\text{mm,1}(V_\text{DC})$ are known from the
	varactor $C$--$V$ characteristic (Sec.~\ref{subsec:circuit_model}).
	
	\subsection{TX Operation: Multiharmonic Generation and Leaky-Wave Radiation}
	\label{subsec:tx}
	
	A baseband signal at $\omega_\text{tx}$ launched at the TX port ($x=0$)
	transfers power into the harmonic ladder $\omega_{n,m} = n\Omega +
	m\omega_\text{tx}$ as it propagates along $x$, governed by
	Eqs.~\eqref{eq:coupled_I}--\eqref{eq:coupled_V}. Each harmonic $(n,m)$
	carries guided wavenumber $\beta_{n,m} = n k_\text{m} + m\beta_\text{tx}$. Whether that
	harmonic radiates into free space or remains bound to the structure is
	determined by comparing $\beta_{n,m}$ with the free-space wavenumber
	$k_{0,n,m} = \omega_{n,m}/c$:
	
	\begin{equation}\label{eq:fast_slow_wave}
		\begin{split}
			|\beta_{n,m}| < k_{0,n,m}&  \qquad \text{fast wave, radiates at }\\
			&\sin\theta_{n,m} = \frac{\beta_{n,m}}{k_{0,n,m}}
			= \frac{n k_\text{m} + m\beta_\text{tx}}{\omega_{n,m}/c},\\
			|\beta_{n,m}| > k_{0,n,m}& \qquad \text{slow wave, remains guided.}
		\end{split}
	\end{equation}

	Equation~\eqref{eq:fast_slow_wave} gives the harmonic-dependent radiation angle for components satisfying the fast-wave condition. For the measured first sidebands, $(n,m)=(1,+1)$ and $(1,-1)$. Since $\omega_\text{tx}/(2\pi)\ll \Omega/(2\pi)$ and $\beta_\text{tx}\ll k_\text{m}$, the two products have nearly equal tangential and free-space wavenumbers:
\begin{equation}
\beta_{1,+1}\approx\beta_{1,-1}\approx k_\text{m},
\qquad
k_{0,1,+1}\approx k_{0,1,-1}.
\label{eq:sideband_angle_approx}
\end{equation}
Consequently, Eq.~\eqref{eq:fast_slow_wave} predicts $\theta_{1,+1}\approx\theta_{1,-1}$, so the first sidebands should occupy nearly the same angular region. This represents harmonic-dependent directional radiation, not independent steering of the two sidebands. Harmonics satisfying the slow-wave condition remain guided and continue toward the RX port. Because $\beta_{n,m}$ depends on
	$\chi_\text{ee,0}\chi_\text{mm,0}$ through Eq.~\eqref{eq:beta0}, and the
	per-harmonic coupling strength depends on the modulation depths
	$\chi_\text{ee,1}/\chi_\text{ee,0}$ and $\chi_\text{mm,1}/\chi_\text{mm,0}$ through
	Eqs.~\eqref{eq:coupled_I}--\eqref{eq:coupled_V}, which are set by the bias point $V_\text{DC}$ and the
	modulation amplitude $V_\text{mod}$ through the varactor $C$--$V$ law (Sec.~\ref{subsec:circuit_model}), so that adjusting the bias redistributes radiated power among the $(n,m)$ channels without
	mechanical steering or phase shifters, giving the bias-reconfigurable
	harmonic spectrum reported in Sec.~\ref{sec:results}.

	\subsection{RX Operation: Phase Matching and Directional Port Routing}
	\label{subsec:rx}
	
	In receive mode, a free-space wave at frequency $\omega_\text{inc}$ incident at angle $\theta_\text{inc}$ illuminates the array and couples
	into the guided line through the time-reversed counterpart of
	Eq.~\eqref{eq:fast_slow_wave}. Efficient coupling into a target guided
	channel $\omega_\text{RX}$ at the RX port requires simultaneous frequency
	and momentum matching,
\begin{subequations}
	\begin{align}
		\omega_\text{inc} &= \omega_\text{RX} + n^{*}\Omega,
		\label{eq:rx_freq_match} \\[2pt]
		\frac{\omega_\text{inc}}{c}\sin\theta_\text{inc} &= \beta_\text{RX} + n^{*} k_\text{m},
		\label{eq:rx_mom_match}
	\end{align}
\end{subequations}
for some integer $n^{*}$. Equations~\eqref{eq:rx_freq_match}--\eqref{eq:rx_mom_match}
	are two conditions on the two free incidence parameters
	$(\omega_\text{inc},\theta_\text{inc})$, so for a given incidence condition only
	one channel $n^{*}$ phase-matches strongly. Here
	$\beta_\text{RX}\equiv\beta_0(\omega_\text{RX})$ is the unmodulated guided wavenumber of
	Eq.~\eqref{eq:beta0} at the RX frequency $\omega_\text{RX}=\omega_\text{inc}-n^{*}\Omega$.
	To quantify the remaining channels, note that the $n$th modulation
	harmonic, which carries $(n\Omega,\,n k_\text{m})$, couples the incident wave
	$(\omega_\text{inc},\,k_x)$, with $k_x=(\omega_\text{inc}/c)\sin\theta_\text{inc}$, to a
	guided component at frequency $\omega_\text{inc}-n\Omega$. Energy conservation fixes that
	frequency and momentum conservation requires the wavenumber $k_x-n k_\text{m}$, whereas
	the structure supports at that frequency only the guided wavenumber
	$\beta_0(\omega_\text{inc}-n\Omega)$. The phase mismatch of order $n$ is therefore
	\begin{equation}
		\Delta\beta_n = \frac{\omega_\text{inc}}{c}\sin\theta_\text{inc}
		- n k_\text{m} - \beta_0\!\left(\omega_\text{inc}-n\Omega\right),
		\label{eq:phase_mismatch}
	\end{equation}
	which vanishes for $n=n^{*}$ by Eqs.~\eqref{eq:rx_freq_match}--\eqref{eq:rx_mom_match}
	and is generically nonzero for every other order, $n\neq n^{*}$. The mismatch
	is thus taken between the momentum supplied by the incident field and the modulation,
	$k_x-nk_\text{m}$, and the guided wavenumber at the shifted frequency.
	Where the guided dispersion is locally linear, $\beta_0(\omega)\approx\omega/v_\text{g}$
	with group velocity $v_\text{g}$, subtracting Eq.~\eqref{eq:phase_mismatch} evaluated at $n^{*}$ gives the
	explicit form
	\begin{equation}
		\Delta\beta_n \approx (n-n^{*})\left(\frac{\Omega}{v_\text{g}}-k_\text{m}\right)
		= (n-n^{*})\,k_\text{m}\left(\frac{v_\text{m}}{v_\text{g}}-1\right),
		\label{eq:phase_mismatch_lin}
	\end{equation}
	where $v_\text{m}=\Omega/k_\text{m}$, so the mismatch grows linearly with the distance $|n-n^{*}|$ from the phase-matched
	order and vanishes for all orders only in the exactly synchronous case $v_\text{m}=v_\text{g}$, so the
	suppression of the sidebands at the RX port requires $|\Omega/v_\text{g}-k_\text{m}|L\gtrsim\pi$. For $\Omega$ comparable to $\omega_\text{inc}$ the dispersion is not linear over the
	span $n\Omega$, and Eq.~\eqref{eq:phase_mismatch} should be used directly.
	To first order in the (weak) modulation depth, the coupled power of order $n$
	over the TX--RX interaction length $L$ follows the standard
	coupled-mode phase-mismatch factor
	\begin{equation}
		P_n \;\propto\; \operatorname{sinc}^2\!\left(\frac{\Delta\beta_n L}{2}\right),
		\qquad
		\operatorname{sinc}(u) \equiv \frac{\sin u}{u}.
		\label{eq:sinc_suppression}
	\end{equation}
	
	Equation~\eqref{eq:sinc_suppression} is the origin of the weak
	sideband levels observed at the RX port: suppression of $n \ne n^{*}$
	follows directly from the coupled-mode phase mismatch, analogous to
	quasi-phase-matched conversion in nonlinear optics, rather than from
	an assumed filtering step.
	
	Nonreciprocity follows from the one-way space-time coherency between the incoming
	wave and the modulation~\cite{Taravati_PRB_SB_2017}. The bias
	$\cos(\Omega t - k_\text{m}x)$ propagates in $+x$ only. A guided wave travelling toward the RX
	port propagates along the modulation, so in the frame moving with the modulation it
	sees an almost stationary pattern and its interaction with the modulation accumulates
	coherently along the aperture. A guided wave travelling toward the TX port propagates
	against the modulation, sees a rapidly oscillating pattern, and its interaction averages out.
	This is quantified by generalizing Eq.~\eqref{eq:phase_mismatch} to a guided wave of direction
	$s=\pm1$ ($s=+1$ toward RX, $s=-1$ toward TX),
	\begin{equation}
		\begin{split}
			\Delta\beta_n^{(s)} =& \frac{\omega_\text{inc}}{c}\sin\theta_\text{inc}
			- n k_\text{m} - s\,\beta_0\!\left(\omega_\text{inc}-n\Omega\right) \\
			&\approx \Big[k_x - s\frac{\omega_\text{inc}}{v_\text{g}}\Big]
			+ n\Big(s\frac{\Omega}{v_\text{g}}-k_\text{m}\Big),
			\label{eq:phase_mismatch_dir}
		\end{split}
	\end{equation}
	where the last form holds for locally linear dispersion. The incident wave is matched toward RX
	at $n^{*}$, $\Delta\beta_{n^{*}}^{(+)}=0$, and the order-to-order change of the mismatch is
	$|\Omega/v_\text{g}-k_\text{m}|$, which is small when the wave and the modulation travel at
	comparable velocities. Toward TX the same incident wave $(\omega_\text{inc},k_x)$ is
	mismatched at every order, with the bracketed term of Eq.~\eqref{eq:phase_mismatch_dir} now of the order of
	$2\beta_0$, and the order-to-order change is $\Omega/v_\text{g}+k_\text{m}$, larger than toward RX. The incident power is therefore routed to the
	RX port and not to the TX port. Reversing the direction of the modulation, $k_\text{m}\to-k_\text{m}$,
	exchanges the roles of the two ports, which is the origin of TX/RX isolation without a
	circulator. It is quantified experimentally by the port-to-port comparison for the same
	illumination in Sec.~\ref{sec:results}.
	
	\subsection{Summary of Distinguishing Features}\label{subsec:summary}
The proposed structure combines three features within one aperture:
\begin{itemize}
\item an omega-shaped bianisotropic unit cell with voltage-dependent loads in both the predominantly electric and ring-associated magnetic current paths;
\item a common traveling-wave bias that makes the composite response periodic in space and time; and
\item access to radiating and guided members of the same harmonic ladder, enabling measured polychromatic radiation and preferential receive-port coupling.
\end{itemize}
These results demonstrate functions relevant to shared-aperture transceivers. They do not constitute a demonstration of simultaneous full-duplex communication or a complete forward-reverse reciprocity measurement.

\subsection{Equivalent-Circuit Model of the Periodic Varactor-Loaded Unit Cell}
	\label{subsec:circuit_model}
	
	This subsection relates the physical varactor-loaded unit cell to the
	abstract susceptibilities $\chi_\text{ee}(x,t)$, $\chi_\text{mm}(x,t)$ of
	Sec.~\ref{subsec:unitcell_response} and to the discrete lattice of the fabricated array. We first describe
	the cell by a distributed-element circuit and then reduce it to the dipole and ring resonators
	whose varactor capacitances carry the modulation.
	
	\begin{figure}
		\begin{center}
			\includegraphics[width=0.8\columnwidth]{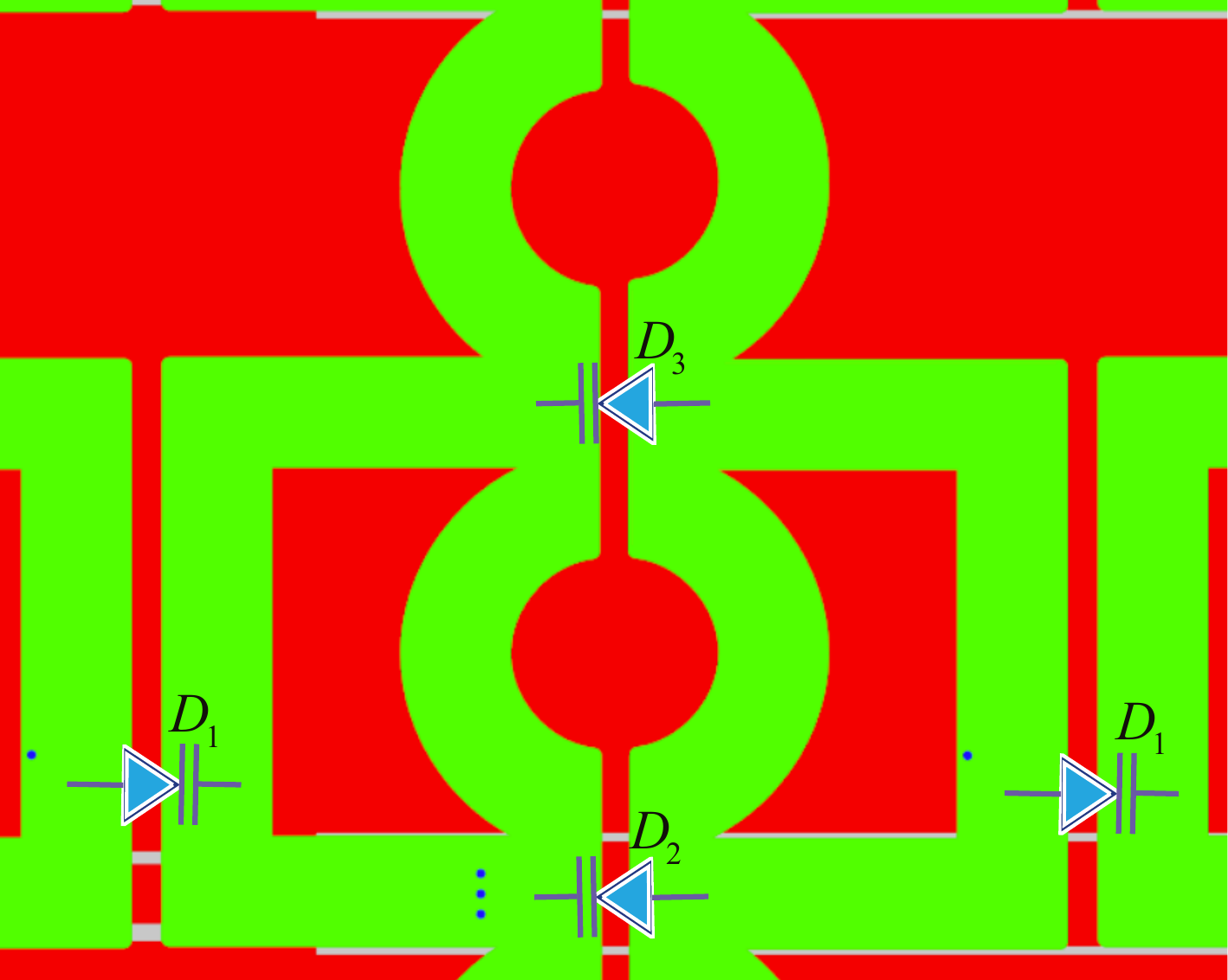}
			\caption{Schematic of the proposed space-time-periodic bianisotropic metasurface with an omega-shaped unit cell. (a)~Unit-cell, showing the dipole varactor $\text{D}_1$ and split-ring varactors $\text{D}_2$, $\text{D}_3$. (b)~Metasurface architecture showing the front (radiating) layer and back (bias/modulation) layer.}
			\label{Fig:Scha}
		\end{center}
	\end{figure}
	
	\begin{figure*}
		\begin{center}
			\includegraphics[width=2\columnwidth]{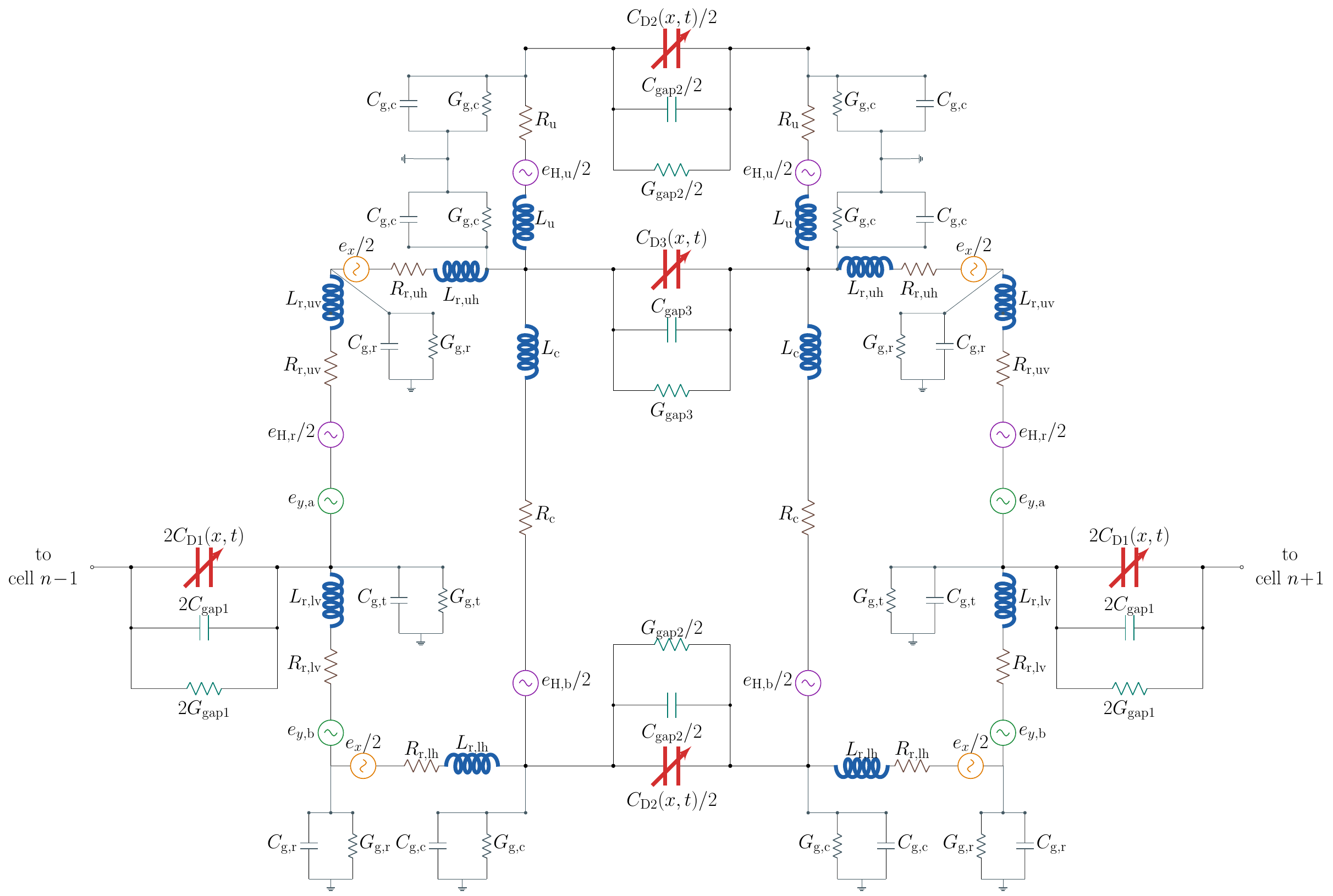}
			\caption{Lumped RLGC equivalent circuit of one bianisotropic metasurface unit cell.
				Inductors (blue) and resistors (brown) represent the metallic segments of the
				rectangular frame ($L_{\mathrm{r},\cdot}$, $R_{\mathrm{r},\cdot}$; subscripts
				u/l for the upper/lower half and h/v for horizontal/vertical bars), of the upper
				circular ring ($L_{\mathrm{u}}$, $R_{\mathrm{u}}$) and of the lower circular ring
				($L_{\mathrm{c}}$, $R_{\mathrm{c}}$). The varactors $D_1$--$D_3$ (red) are
				time- and space-dependent capacitances $C_{D_j}(x,t)$, each in parallel with its
				fixed gap capacitance $C_{\mathrm{gap}}$ and conductance $G_{\mathrm{gap}}$ (green).
				$D_1$ bridges the slit to the neighboring cells, so each cell carries two halves
				of value $2C_{D_1}$; $D_2$ is shared with the cells above and below, so each cell
				carries $C_{D_2}/2$ at its top and bottom necks. The grey pairs
				$C_{\mathrm{g},x}$, $G_{\mathrm{g},x}$ are the ground-plane shunt elements at the
				rectangular-ring ($x=\mathrm{r}$), circular-ring ($x=\mathrm{c}$) and $D_1$-tap
				($x=\mathrm{t}$) nodes. Sources: $e_x/2$ (horizontal dipole, one per horizontal
				bar), $e_{y,\mathrm{a}}$, $e_{y,\mathrm{b}}$ (vertical dipole, split by the $D_1$
				tap) and $e_{\mathrm{H,u}}/2$, $e_{\mathrm{H,r}}/2$ (loop emfs of the upper and
				lower rings). The rings are not driven by $H_z$ at normal incidence; they are driven through the shared nodes by the frame sources $e_x/2$ and $e_y$.}
			\label{Fig:Equi}
		\end{center}
	\end{figure*}

	\subsubsection{Distributed-Element Circuit of the Unit Cell}
	\label{subsubsec:full_circuit}
	
	Fig.~\ref{Fig:Scha} depicts the proposed unit cell and Fig.~\ref{Fig:Equi} gives the equivalent circuit of the unit cell, from which the reduced
	dipole and ring resonators of the following subsubsections are
	obtained. The cell contains three metallic rings, an upper circular
	ring, a lower circular ring and a rectangular frame enclosing the lower
	ring, joined by the varactors $D_3$ (neck between the two circles),
	$D_2$ (bottom neck, shared with the next cell along $y$) and $D_1$
	(slit between the frames of adjacent cells along $x$). Topologically the
	cell is a two-rail ladder along $y$: the rungs are $D_3$ and the two
	half-$D_2$ branches, and the rails carry the segment inductances. Each metallic segment is modeled by a series inductance and resistance,
	$L_s$ and $R_s$, as in the standard lumped description of split-ring
	structures. Each varactor is shunted by the capacitance and
	conductance of the physical gap it bridges, so the total gap
	admittance is
	$Y_{D_j} = G_{\mathrm{gap}j} + j\omega\,[C_{\mathrm{gap}j} + C_{D_j}(x,t)]$.
	Because $D_1$ is a series element of the $x$-chain, each cell owns two
	halves of it, each of value $2C_{D_1}$ with $2C_{\mathrm{gap}1}$ and
	$2G_{\mathrm{gap}1}$ in parallel. Because $D_2$ is a shunt rung shared
	with the neighbor along $y$, each cell owns $C_{D_2}/2$,
	$C_{\mathrm{gap}2}/2$ and $G_{\mathrm{gap}2}/2$ at its top and bottom
	necks, so that the shared gap is recovered when two cells are joined.
	
	The conductors also couple to the ground plane through the substrate.
	This is represented by shunt pairs $C_{\mathrm{g},x}$, $G_{\mathrm{g},x}$ at
	the nodes, with $x=\mathrm{r}$ for the rectangular frame, $x=\mathrm{c}$ for
	the circular rings and $x=\mathrm{t}$ for the short taps at $D_1$. Each pair
	is the substrate loading of the segments adjoining the node,
	\begin{equation}
		C_{\mathrm{g},x} = c_{\mathrm{g}}\,A_x,
		\qquad
		G_{\mathrm{g},x} = \omega\,C_{\mathrm{g},x}\tan\delta,
		\label{eq:cg_gg}
	\end{equation}
	where $c_{\mathrm{g}}$ is the capacitance per unit area, $A_x$ the area of
	the adjoining conductor and $\tan\delta$ the substrate loss tangent, so the
	three classes differ in magnitude but share the ratio
	$G_{\mathrm{g},x}/C_{\mathrm{g},x}$. Differential modes (the ring and
	the horizontal dipole) see $C_{\mathrm{gap}}+C_{\mathrm{g}}/2$ across
	the slit, whereas the common mode driven by the vertical field
	component sees the full $C_{\mathrm{g}}$.
	
	The incident field enters through distributed electromotive forces. The
	horizontal dipole is driven by $e_x/2$ in each horizontal frame bar,
	and the vertical dipole by $e_{y,\mathrm{a}}$ and $e_{y,\mathrm{b}}$ in
	the vertical frame sides, which together make up the vertical-dipole emf
	$e_y$ divided by the $D_1$ tap. The circular rings respond only to the
	normal magnetic field through the loop emf
	$e_{\mathrm{H}}=j\omega\mu_0 A\,H_z$, placed in the arms of each ring
	as $e_{\mathrm{H,u}}/2$ and $e_{\mathrm{H,r}}/2$. For a plane wave
	$H_z=(k_xE_y-k_yE_x)/(\omega\mu_0)$, so the loop emfs of the rings vanish at
	normal incidence. The rings are then driven electrically: they share nodes
	with the frame, so the sources $e_x/2$ and $e_y$ of the dipoles drive ring
	currents through the shared nodes (bianisotropic coupling). At oblique
	incidence the loop emfs add to this drive.
	
	Finally, the varactor capacitances carry the traveling-wave bias,
	$C_{D_j}(x,t)$ with $V_j(x,t)=V_{\mathrm{DC}}+V_{\mathrm{mod}}\cos(\Omega t-k_m x)$, which
	makes the circuit time-varying and periodic in $x$. If the bias is applied
	through one island of the frame, the left and right $D_1$ halves of a cell
	see modulation phases that differ by $\Delta\varphi$ set by the
	bias-line spacing. The element values are effective parameters extracted around the
	fundamental resonances, in the narrow band of interest; as for lumped models of split-ring resonators, the
	segments are electrically short only approximately, so the circuit is
	accurate near these resonances and not meant to reproduce the full
	wideband response. Near resonance the network collapses onto the two
	resonators used below: the dipole branch with $L_e$ and
	$C_e(t)=C_{\mathrm{gap}}+C_{D_1}(t)$, and the ring branch with $L_m$ and
	$C_m(t)$ set by $D_2$ and $D_3$ together with their gap capacitances.

	\subsubsection{Varactor $C$--$V$ Characteristic}
	
	Each diode $\text{D}_j$ ($j=1,2,3$) is modeled by the standard
	hyperabrupt-junction relation
	\begin{equation}
		C_{\text{D}_j}\big(V_j(t)\big) =
		\frac{C_{j0}}{\left(1 + V_j(t)/V_\text{bi}\right)^{M_j}},
		\label{eq:cv_curve}
	\end{equation}
	where $C_{j0}$ is the zero-bias capacitance, $V_\text{bi}$ the built-in
	potential, and $M_j$ the grading coefficient, all obtained from the
	varactor datasheet.
	
	\subsubsection{Discrete Traveling-Wave Bias}
	The power-divider/delay network applies a fixed phase increment per
	cell rather than a continuous $k_\text{m} x$. For cell $q$ located at
	$x_q = q\,p$ (with $p$ the lattice period),
	\begin{equation}
		V_j(x_q,t) = V_\text{DC} + V_\text{mod}\cos\!\big(\Omega t - q\,\Delta\phi\big),
		\label{eq:discrete_bias}
	\end{equation}
	where $\Delta\phi = k_\text{m}\,p$. 
	
	\subsubsection{Linearized Varactor Modulation}
	For $V_\text{mod} \ll V_\text{bi}+V_\text{DC}$, with $V_j$ the reverse bias, Eq.~\eqref{eq:cv_curve} may be linearized about the bias point:
	\begin{subequations}
		\begin{equation}\label{eq:cv_linearized}
			C_{\text{D}_j}(t) \approx C_{j,0} + C_{j,1}\cos(\Omega t - k_\text{m}x),
		\end{equation}
		\begin{equation}
			C_{j,1} = S_j\, V_\text{mod},
		\end{equation}
		\begin{equation}	\label{eq:sensitivity}
			S_j \equiv \left.\frac{d C_{\text{D}_j}}{dV}\right|_{V_\text{DC}}
			= -\frac{M_j\, C_{j0}}{V_\text{bi}}
			\left(1 + \frac{V_\text{DC}}{V_\text{bi}}\right)^{-(M_j+1)}.
		\end{equation}
	\end{subequations}
	The slope magnitude $|S_j|$ decreases monotonically with the reverse bias, and so does the
	normalized modulation index
	\begin{equation}
		\mu_j \equiv \frac{|C_{j,1}|}{C_{j,0}} = \frac{M_j\,V_\text{mod}}{V_\text{bi}+V_\text{DC}},
		\label{eq:mod_index}
	\end{equation}
	with $C_{j,0}=C_{j0}(1+V_\text{DC}/V_\text{bi})^{-M_j}$. Hence, at fixed $V_\text{mod}$, a larger
	$V_\text{DC}$ gives a smaller capacitance swing, not a larger one. When $V_\text{mod}$ is not small
	against $V_\text{bi}+V_\text{DC}$, or when the swing approaches the forward-conduction limit,
	$V_\text{DC}-V_\text{mod}\to0$, the linearization is not valid and the capacitance waveform
	$C_{\text{D}_j}(t)=C_{\text{D}_j}\big(V_\text{DC}+V_\text{mod}\cos\Omega t\big)$ must be expanded
	in its full Fourier series, whose coefficients $C_{j,k}=\frac{\Omega}{\pi}\int_0^{2\pi/\Omega}
	C_{\text{D}_j}(t)\cos(k\Omega t)\,dt$ ($k\ge1$) also fix the strength of the higher-order terms.
	\subsubsection{Dipole (Electric) Branch}
	The loaded dipole arm is modeled as a series $RLC$ resonator with
	self-inductance $L_e$, damping rate $\gamma_e = R_e/L_e$, and total
	gap capacitance $C_e(t) = C_\text{gap} + C_{\text{D}_1}(t)$, where $C_\text{gap}$ is the
	intrinsic (unloaded) gap capacitance in parallel with $\text{D}_1$. Its
	instantaneous resonance frequency is
	\begin{equation}
		\omega_e(t) = \frac{1}{\sqrt{L_e\, C_e(t)}}.
		\label{eq:omega_e}
	\end{equation}
	Homogenizing over the areal unit-cell density $N = 1/(p\,w)$ (with $w$
	the transverse cell dimension) gives the electric surface
	susceptibility used in
	Sec.~\ref{subsec:unitcell_response}:
	\begin{equation}
		\chi_\text{ee}(\omega,t) =
		\frac{N\, p_e^{\,2}\, \omega^2 / \epsilon_0}
		{\omega_e(t)^2 - \omega^2 - i\,\omega\,\gamma_e},
		\label{eq:chi_ee_lorentz}
	\end{equation}
	where $p_e$ is the dipole's effective moment-per-field, set by the arm
	length and $L_e$ via standard dipole-antenna theory.
	
	\subsubsection{Ring (Magnetic) Branch}
	The split ring is modeled analogously as a parallel resonator with
	self-inductance $L_\text{m}$, damping rate $\gamma_\text{m} = R_\text{m}/L_\text{m}$, and gap
	capacitance $C_\text{m}(t)$ set by $\text{D}_2$ and $\text{D}_3$. If the two gaps sit in
	series around the loop current path,
\begin{subequations}
	\begin{equation}
		C_\text{m}(t) = \frac{C_{\text{D}_2}(t)\, C_{\text{D}_3}(t)}{C_{\text{D}_2}(t) + C_{\text{D}_3}(t)};
		\label{eq:cm_series}
	\end{equation}
	if instead the two gaps are driven independently rather than sharing
	one loop current path, this becomes
	\begin{equation}
		C_\text{m}(t) = C_{\text{D}_2}(t) + C_{\text{D}_3}(t).
		\label{eq:cm_parallel}
	\end{equation}
	The corresponding resonance and magnetic surface susceptibility are
	\begin{align}
		\omega_\text{m}(t) &= \frac{1}{\sqrt{L_\text{m}\, C_\text{m}(t)}},
		\label{eq:omega_m} \\[4pt]
		\chi_\text{mm}(\omega,t) &=
		\frac{N\, m^2\, \omega^2 / \mu_0}
		{\omega_\text{m}(t)^2 - \omega^2 - i\,\omega\,\gamma_\text{m}},
		\label{eq:chi_mm_lorentz}
	\end{align}
\end{subequations}
	where $m$ is the ring's magnetic moment-per-field, set by the loop
	area.
	
	\subsubsection{Modulation-Depth Ratios}
	For weak modulation, the ratios $\chi_\text{ee,1}/\chi_\text{ee,0}$ and $\chi_\text{mm,1}/\chi_\text{mm,0}$ that
	enter the coupled-mode equations \eqref{eq:coupled_I}--\eqref{eq:coupled_V} follow from the dependence of
	Eqs.~\eqref{eq:chi_ee_lorentz} and \eqref{eq:chi_mm_lorentz} on the capacitance,
	$\chi_\text{ee,1}\approx(\partial\chi_\text{ee}/\partial C_\text{e})\,C_{1,1}$ and
	$\chi_\text{mm,1}\approx(\partial\chi_\text{mm}/\partial C_\text{m})\,C_\text{m,1}$, with $C_{1,1}$ from
	Eq.~\eqref{eq:cv_linearized} and $C_\text{m,1}$ from the combination of Eq.~\eqref{eq:cm_series}
	or \eqref{eq:cm_parallel}. This expansion treats the modulation as slow compared with the resonator response;
	for $\Omega$ comparable to the resonance frequencies, as in the present measurements, it indicates the scaling with the
	varactor parameters ($C_{j0}, V_\text{bi}, M_j, V_\text{DC}, V_\text{mod}$) and not the absolute depth. In this work the
	model is used to identify the mechanisms and the scaling behind the observed effects, namely
	the harmonic ladder, the fast- and slow-wave selection, and the phase-matching and coherency
	conditions of Secs.~\ref{subsec:tx}--\ref{subsec:rx}. It is not fitted to the measured spectra: this would require the extracted values
	of $L_\text{e}$, $L_\text{m}$, $\gamma_\text{e}$, $\gamma_\text{m}$, $p_e$ and $m$ and the resulting
	susceptibilities, which are not part of the present measurements, so the model is not claimed to be
	quantitatively validated here.

	The unit-cell period $p$ is bounded by two conditions: the discrete
	bias sampling of Eq.~\eqref{eq:discrete_bias} must resolve the
	modulation wave without aliasing, and the physical array must avoid
	grating lobes at the highest radiating harmonic $\omega_N$ over the
	intended steering range $\theta_\text{max}$:
	\begin{equation}
		p < \frac{\pi}{k_\text{m}}, \qquad
		p < \frac{\lambda_N}{1 + |\sin\theta_\text{max}|}, \qquad
		\lambda_N = \frac{2\pi c}{\omega_N}.
		\label{eq:lattice_constraint}
	\end{equation}

	\begin{figure}
		\begin{center}
			\includegraphics[width=1\columnwidth]{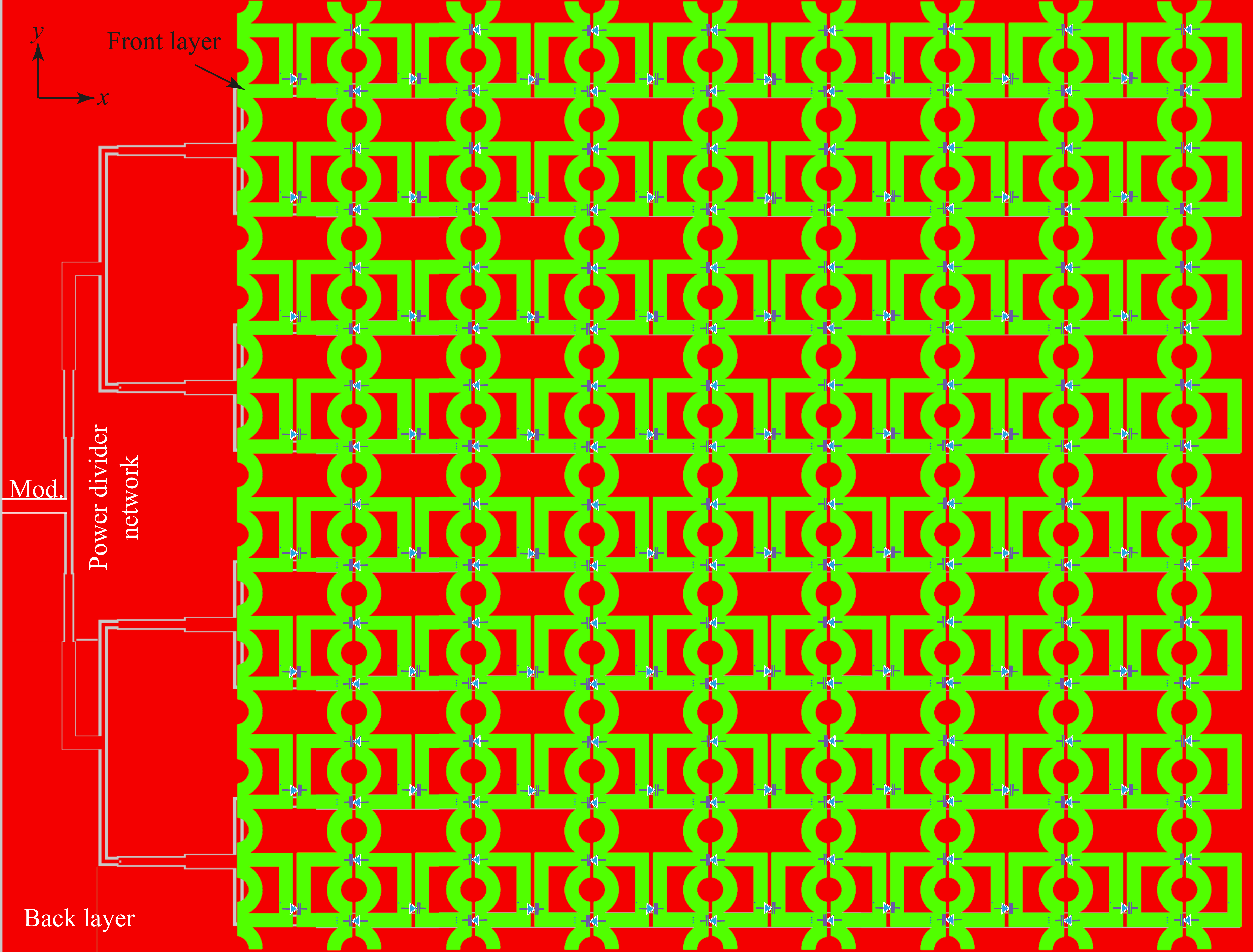}
			\caption{Metasurface architecture showing the front (radiating) layer and back (bias/modulation) layer.}
			\label{Fig:Schb}
		\end{center}
	\end{figure}
	
	Figure~\ref{Fig:Schb} shows the designed space-time-modulated metasurface. The traveling-wave modulation signal is applied on the back layer of the structure, where a corporate power-divider network [Fig.~\ref{Fig:Schb}] distributes the bias waveform to each unit cell with the fixed inter-cell phase increment $\Delta\phi$ of Eq.~\eqref{eq:discrete_bias}; the bias is then transferred from the back layer to the varactors on the front (radiating) layer through metallized via holes at each cell, so that the RF aperture and the DC/modulation distribution network occupy separate layers and do not electromagnetically load one another. Within each unit cell [Fig.~\ref{Fig:Scha}], the dipole-plus-split-ring structure is divided into two electrically distinct halves, left and right. The left half, containing the varactor cathodes, is connected to the positive DC bias node ($V_\text{DC}+V_\text{mod}\cos(\Omega t - k_\text{m} x)$); the right half, containing the varactor anodes, is connected to the zero-DC (ground) node. This left/right split biasing scheme reverse-biases $\text{D}_1$, $\text{D}_2$, and $\text{D}_3$ across each cell using a single traveling-wave bias line per cell, while keeping the RF ports of the dipole and split ring isolated from the DC path via the intrinsic diode capacitance -- avoiding the need for separate RF-choke and DC-blocking networks at every varactor.
	
	\section{Results}
	\label{sec:results}
	\begin{figure*}
		\begin{center}
			\subfigure[]{\label{Fig:Ea}
				\includegraphics[width=0.48\columnwidth]{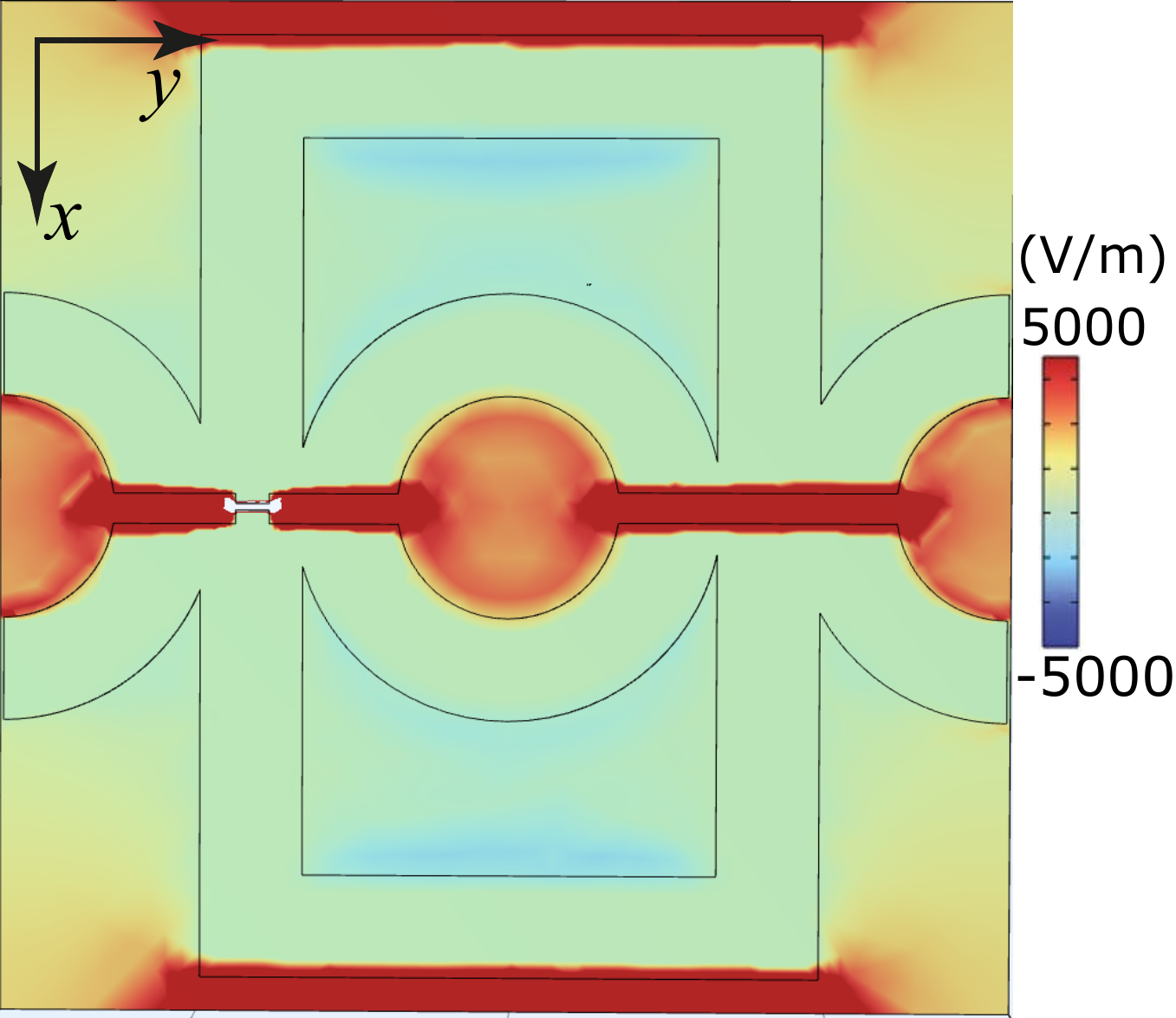}}
			\subfigure[]{\label{Fig:Eb}
				\includegraphics[width=0.48\columnwidth]{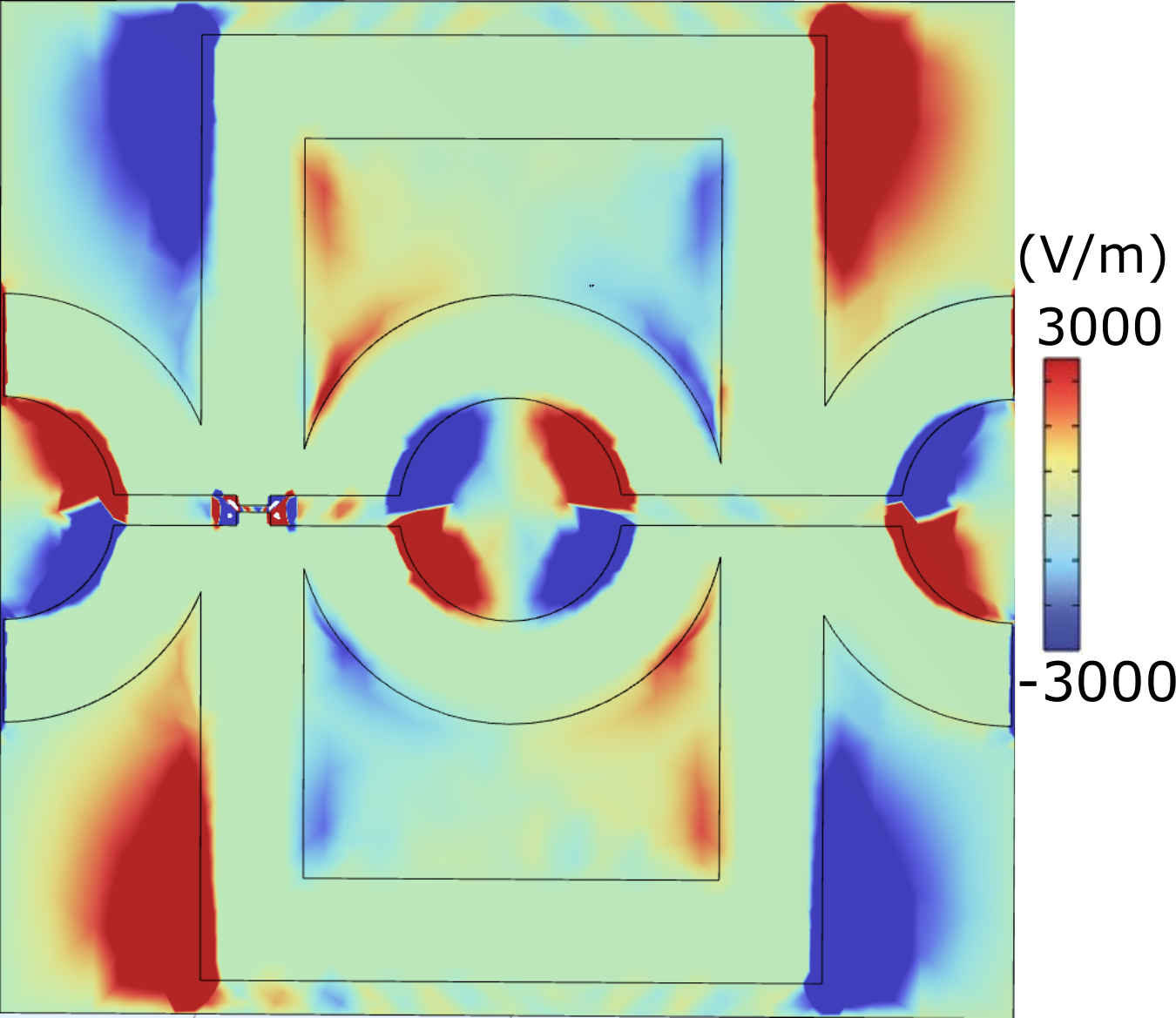}}
			\subfigure[]{\label{Fig:Ec}
				\includegraphics[width=0.48\columnwidth]{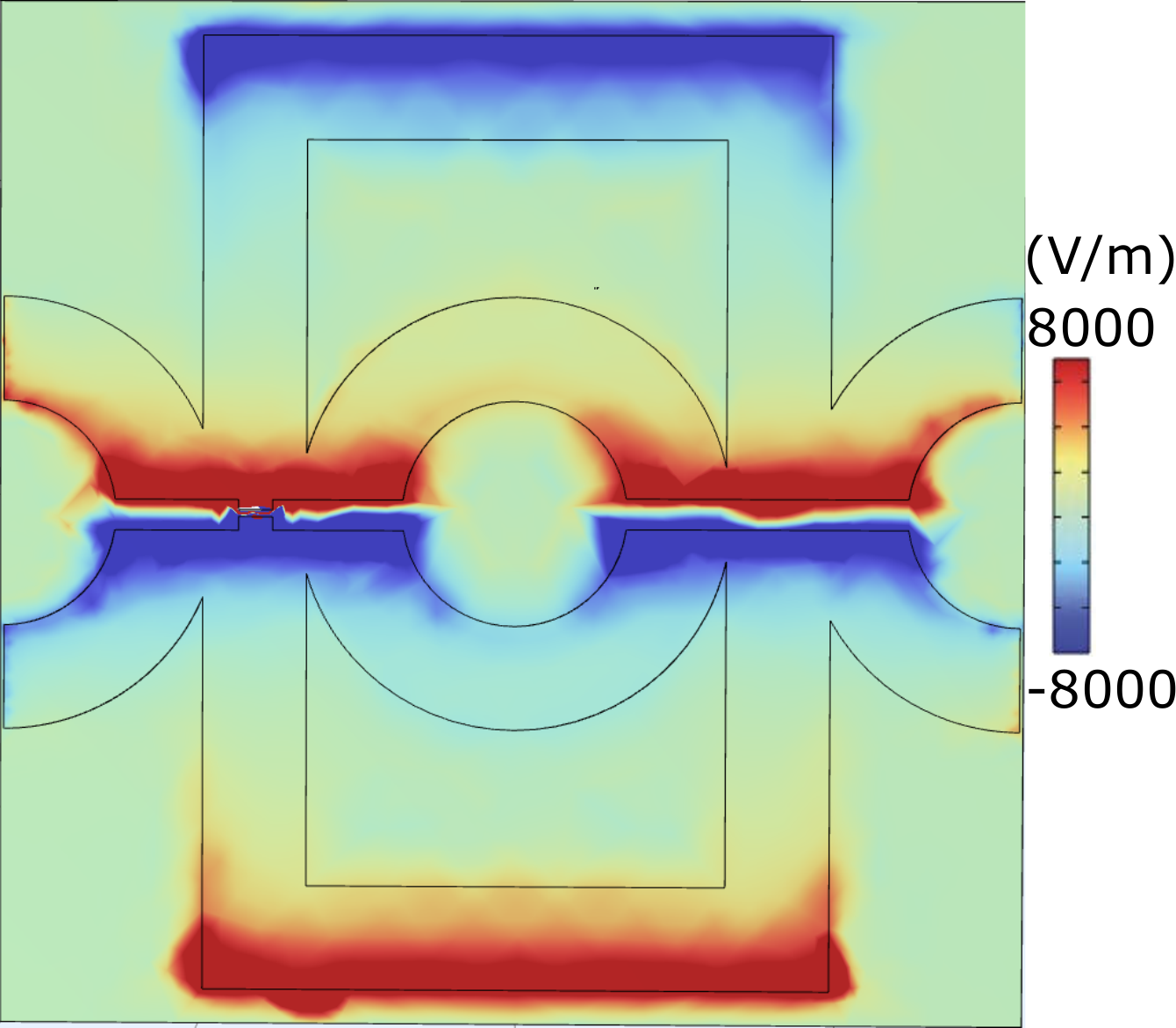}}
			\subfigure[]{\label{Fig:Ed}
				\includegraphics[width=0.5\columnwidth]{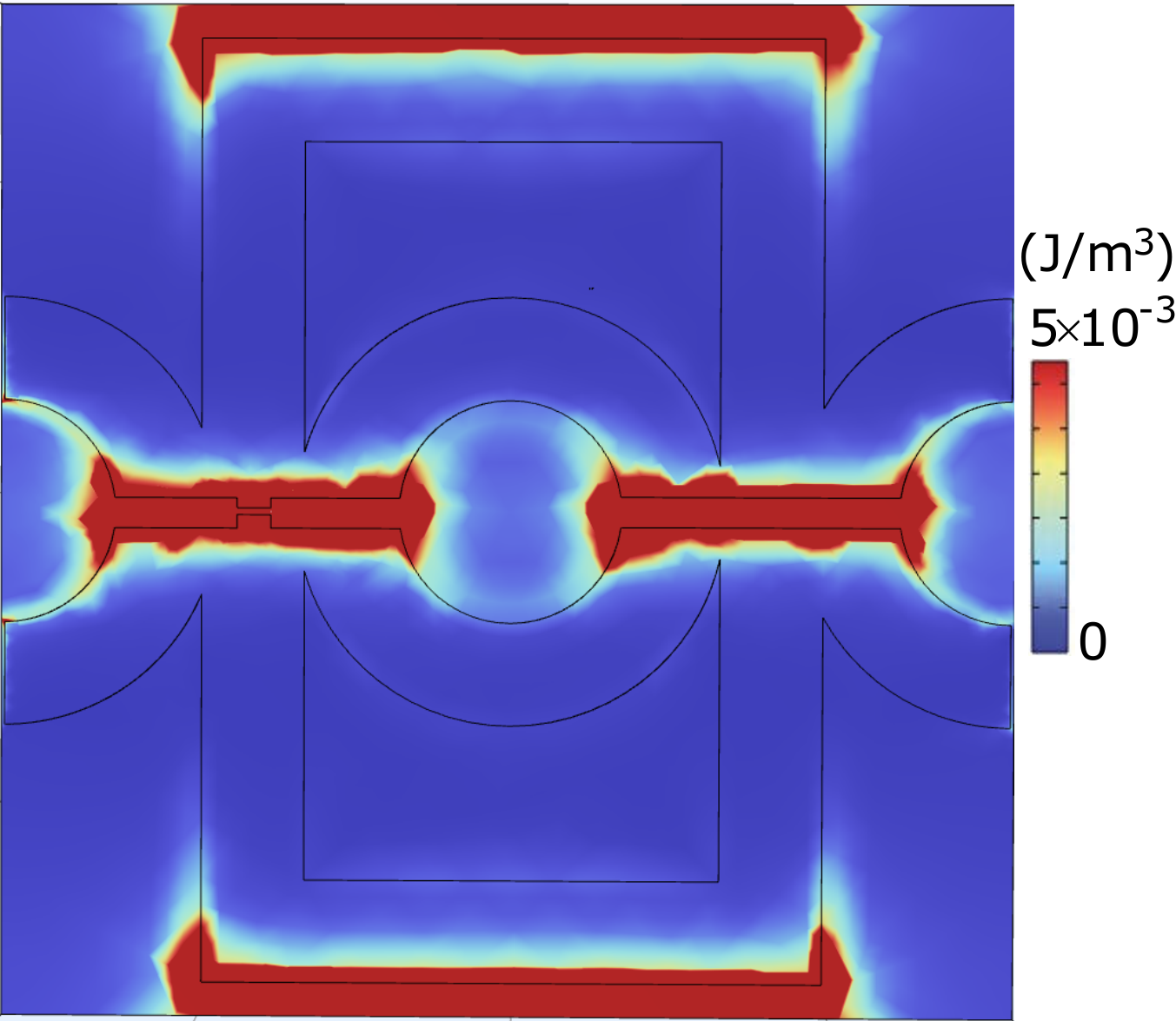}}
			\caption{Full-wave electric response of the unmodulated bianisotropic unit cell at the 1.8-GHz operating frequency, evaluated on a plane immediately above the patterned top metallization. (a)~$E_x$. (b)~$E_y$. (c)~$E_z$. (d)~Time-averaged electric-energy density. The oppositely signed $E_y$ regions inside both circular rings reveal an induced transverse electric response associated with the ring sections.}
			\label{Fig:Esim}
		\end{center}
	\end{figure*}
	
	\begin{figure*}
		\begin{center}
			\subfigure[]{\label{Fig:Ha}
				\includegraphics[width=0.48\columnwidth]{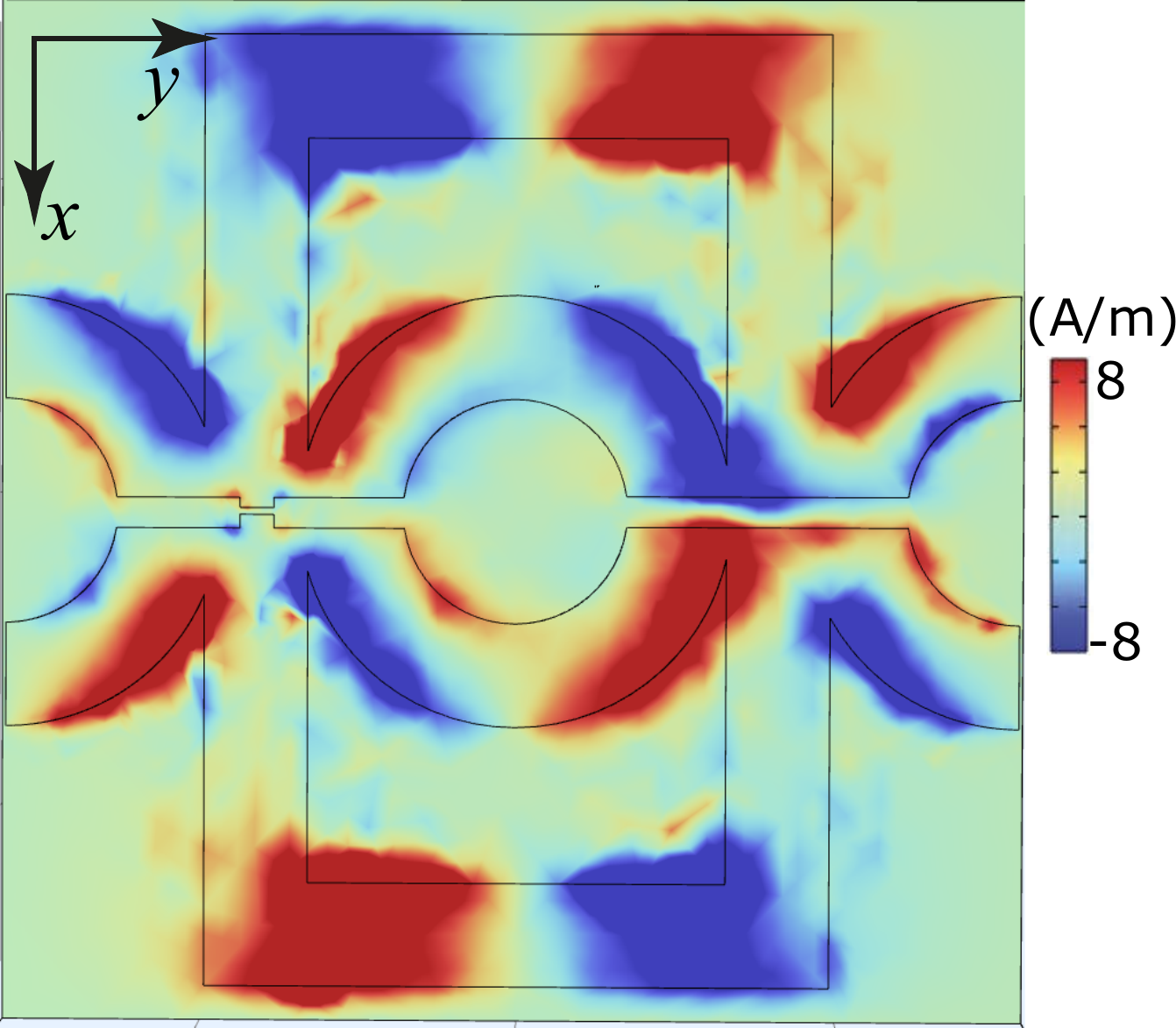}}
			\subfigure[]{\label{Fig:Hb}
				\includegraphics[width=0.48\columnwidth]{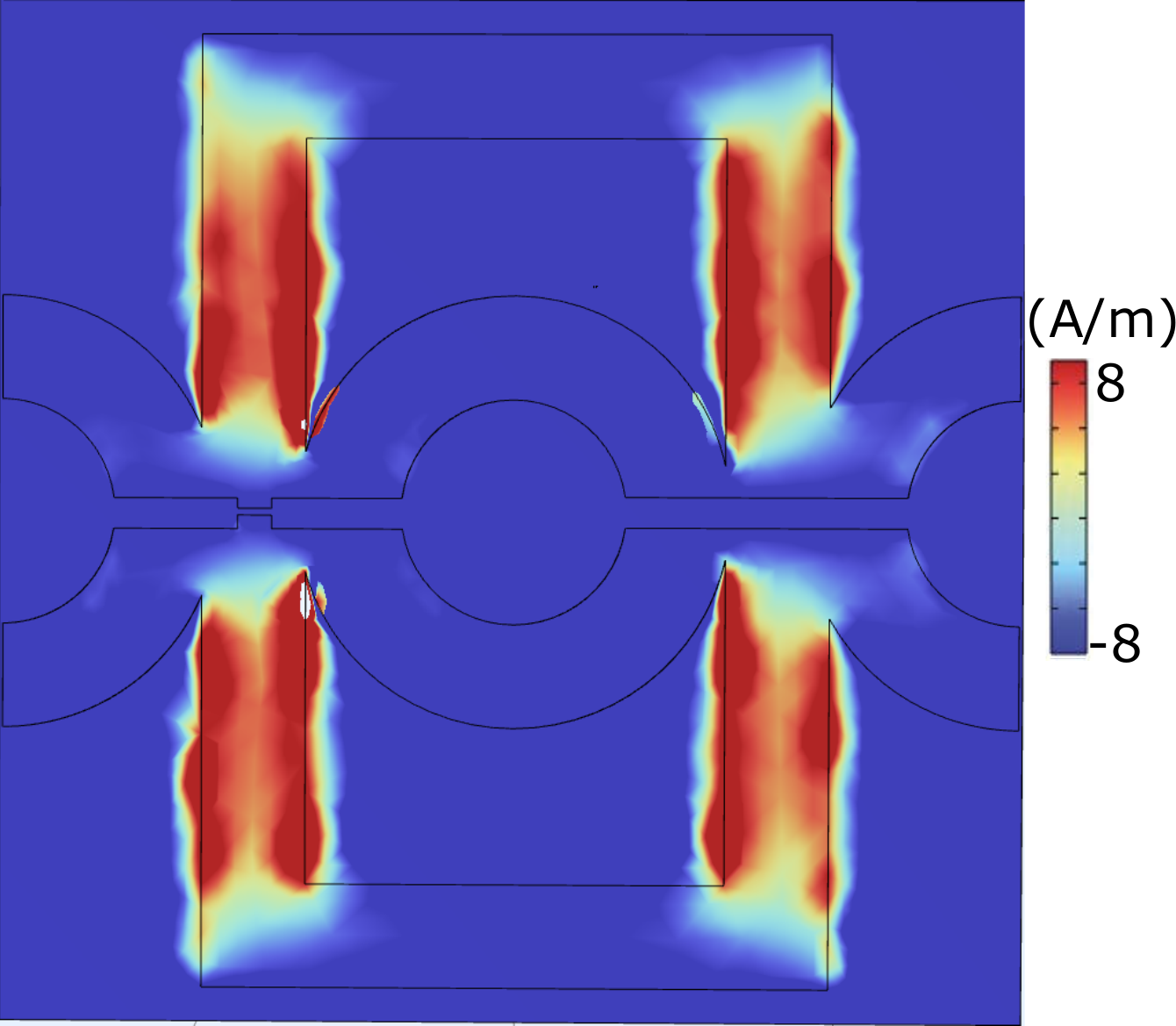}}
			\subfigure[]{\label{Fig:Hc}
				\includegraphics[width=0.48\columnwidth]{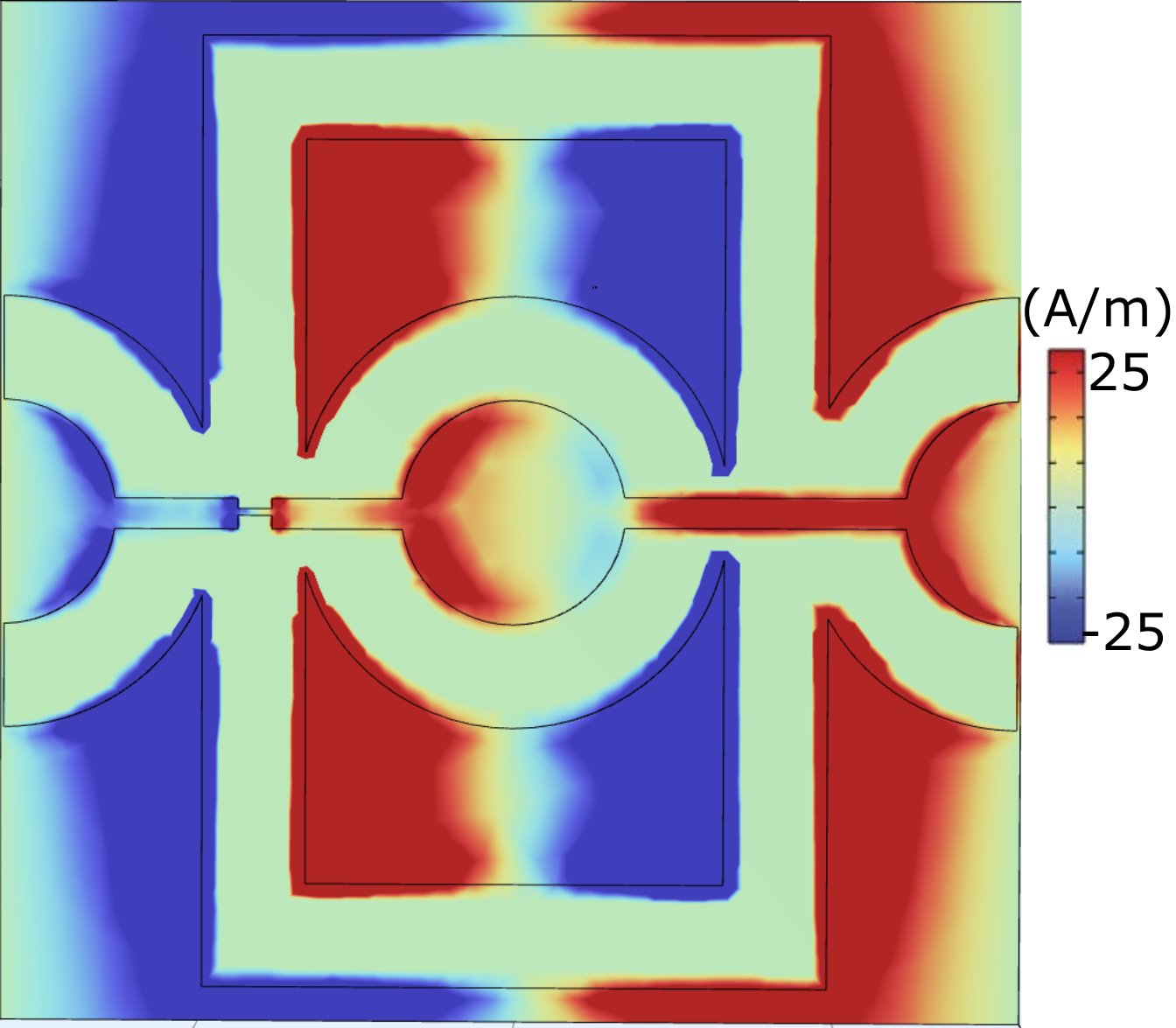}}
			\subfigure[]{\label{Fig:Hd}
				\includegraphics[width=0.5\columnwidth]{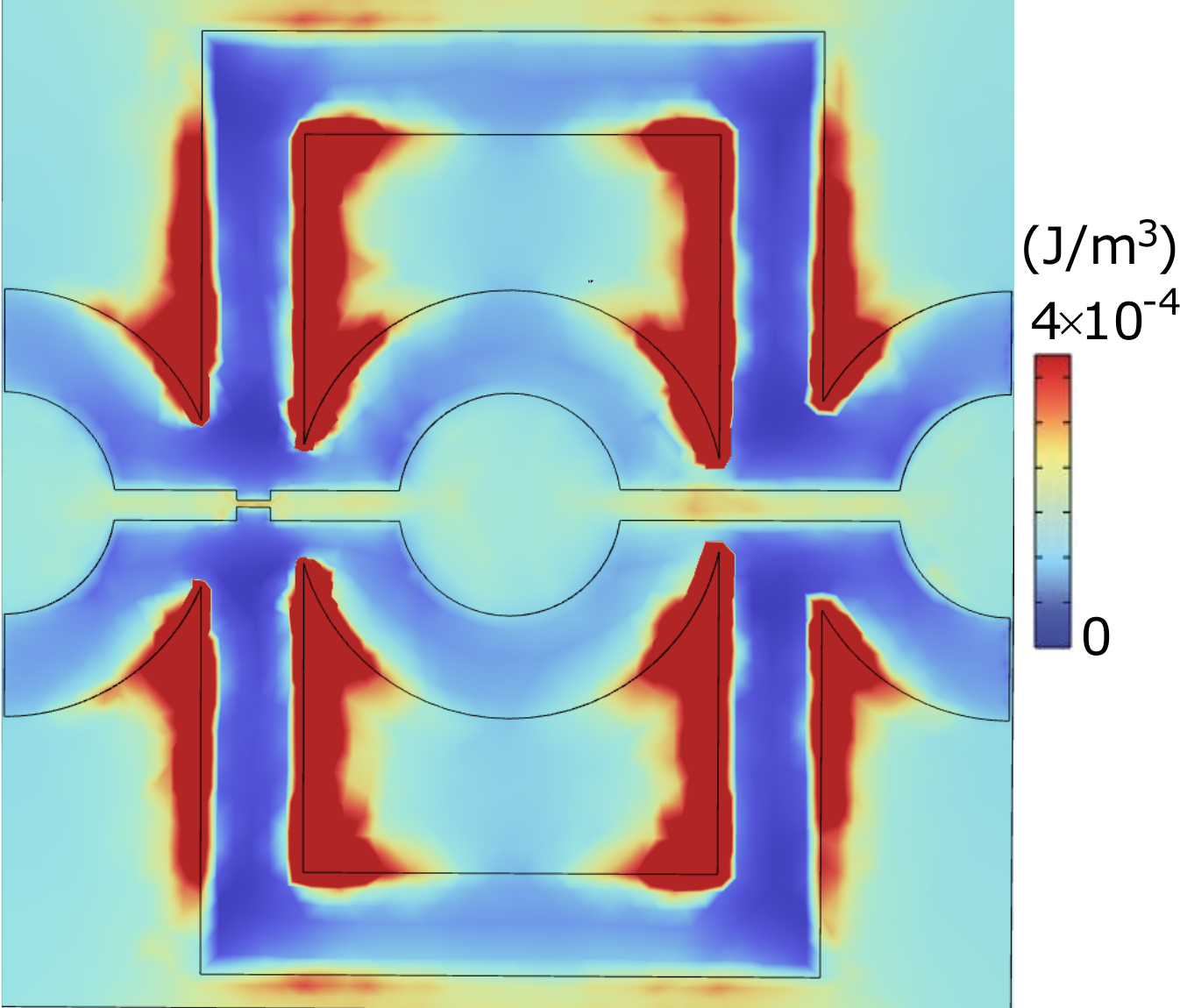}}
			\caption{Full-wave magnetic response of the unmodulated bianisotropic unit cell at the 1.8-GHz operating frequency, evaluated on a plane immediately above the patterned top metallization. (a)~$H_x$. (b)~$H_y$. (c)~$H_z$. (d)~Time-averaged magnetic-energy density. The strong $H_x$ localization within both circular rings, together with the structured $H_y$, $H_z$, and magnetic-energy distributions, shows that the rings are magnetically active at the experimental operating frequency.}
			\label{Fig:Hsim}
		\end{center}
	\end{figure*}

	\subsection{Electric and Magnetic Near-Field Responses at 1.8~GHz}
Full-wave simulations of the unmodulated bianisotropic unit cell were performed with a frequency-domain solver and periodic boundary conditions. The incident plane wave propagates normal to the aperture and is polarized with its electric field along $x$. Figures~\ref{Fig:Esim} and~\ref{Fig:Hsim} are evaluated on a common plane immediately above the patterned top metallization, rather than near the lower substrate boundary. This observation plane samples the near fields directly associated with the conducting meta-atom and therefore reveals substantially stronger local field amplitudes. The absolute color-bar values should consequently be interpreted for this specified plane and should not be compared directly with values obtained at a different height.

The horn antenna employed in the measurements launches a propagating electromagnetic field containing both $\mathbf{E}$ and $\mathbf{H}$; it is not a purely electric induction source. For ideal normal incidence, however, the incident field has no normal magnetic component, $H_z^{\mathrm{inc}}=0$. Any structured $H_z$ field generated around the rings therefore arises from the induced current distribution of the interconnected meta-atom rather than from direct excitation by an incident $H_z$ component.

Figures~\ref{Fig:Ea}--\ref{Fig:Ec} show $E_x$, $E_y$, and $E_z$, respectively. The $E_x$ distribution is dominated by the central horizontal dipolar path and the two circular loading regions, with localized extrema at the narrow connections and capacitive discontinuities. More notably, Fig.~\ref{Fig:Eb} shows pronounced oppositely signed $E_y$ regions inside and around both circular rings. Since the normally incident excitation is polarized along $x$, the incident field has $E_y^{\mathrm{inc}}=0$ in the idealized model. The strong local transverse field is therefore generated by charge redistribution and induced currents in the interconnected omega-shaped conductor rather than being directly imposed by the incident polarization. The red--blue reversal across each ring reflects the signed direction of $E_y$ and identifies a transverse electric polarization associated with the ring sections. The strongest electric component is $E_z$, whose plotted range reaches approximately $\pm8\times10^3$~V/m. It is concentrated along the central horizontal conductor and along the upper and lower horizontal sections of the rectangular frame. The time-averaged electric-energy density in Fig.~\ref{Fig:Ed} reaches the $10^{-3}$-J/m$^3$ scale and is localized along these same conductor edges and capacitive regions. The electric response is therefore distributed over the complete omega-shaped current path rather than being confined to an isolated straight dipole.

Figures~\ref{Fig:Ha}--\ref{Fig:Hc} show the three magnetic-field components on the same top-surface observation plane. Most importantly, the $H_x$ pattern in Fig.~\ref{Fig:Ha} contains strong, oppositely signed magnetic-field regions inside both circular rings. The field changes sign across the upper and lower parts of each loop and extends through the narrow central connections, providing direct spatial evidence that both rings participate in the 1.8-GHz mode. The $H_y$ field in Fig.~\ref{Fig:Hb} is concentrated mainly along the vertical portions of the rectangular frame, with paired positive and negative regions on opposite sides. The $H_z$ component in Fig.~\ref{Fig:Hc} is the strongest plotted magnetic component, reaching approximately $\pm25$~A/m, and exhibits broad, structured regions within the rectangular frame and around the circular openings. Because $H_z^{\mathrm{inc}}=0$ at ideal normal incidence, this normal magnetic field is generated by the electrically induced currents of the composite geometry and is consistent with electric-to-magnetic coupling in the omega-shaped meta-atom.

The time-averaged magnetic-energy density in Fig.~\ref{Fig:Hd} reaches the $10^{-4}$-J/m$^3$ scale and is strongly localized within and around the two circular rings, along the vertical sides of the rectangular frame, and near the narrow interconnections. When considered together, the induced transverse $E_y$ field inside both circular rings, the $H_x$ localization in the same regions, the induced $H_z$ response, and the ring-associated magnetic-energy storage demonstrate that the rings participate in both the electric and magnetic near-field structure at the operating frequency. The spatial coincidence of these secondary electric and magnetic components provides field-level evidence of coupled excitation in the interconnected omega-shaped meta-atom. The 1.8-GHz state should therefore be interpreted as a coupled electric--magnetic mode of the complete cell, not as a purely electric dipole mode.

These near-field maps support the bianisotropic interpretation but do not, by themselves, constitute a retrieval of the magnetoelectric susceptibility tensors. A rigorous quantitative separation of $\overline{\overline{\chi}}_{\mathrm{ee}}$, $\overline{\overline{\chi}}_{\mathrm{mm}}$, $\overline{\overline{\chi}}_{\mathrm{em}}$, and $\overline{\overline{\chi}}_{\mathrm{me}}$ would require complex scattering data under independent illuminations or an equivalent multipolar/current-based retrieval. The present result is a spatially resolved physical observation: an $x$-polarized incident field with $E_y^{\mathrm{inc}}=0$ and $H_z^{\mathrm{inc}}=0$ produces a strong transverse $E_y$ response inside the rings together with pronounced ring-associated $H_x$, $H_z$, and magnetic-energy distributions. This behavior is consistent with the magnetoelectric coupling expected from an omega-type bianisotropic meta-atom, while quantitative confirmation of the cross-coupling terms requires susceptibility or multipolar retrieval.

Figures~\ref{Fig:Esim} and~\ref{Fig:Hsim} characterize the unmodulated cell and therefore do not alone demonstrate temporal modulation. Space-time variation is introduced when the common traveling-wave bias drives $\text{D}_1$, $\text{D}_2$, and $\text{D}_3$. Because these voltage-dependent capacitances load the current paths identified by the electric and magnetic near-field maps, the applied bias varies the collective bianisotropic response in space and time. The subsequent measurements characterize this complete-cell response, without assigning separate experimental modulation depths to the electric, magnetic, and magnetoelectric channels.

\begin{figure}
	\begin{center}
		\subfigure[]{\label{Fig:Phototop}
			\includegraphics[width=1\columnwidth]{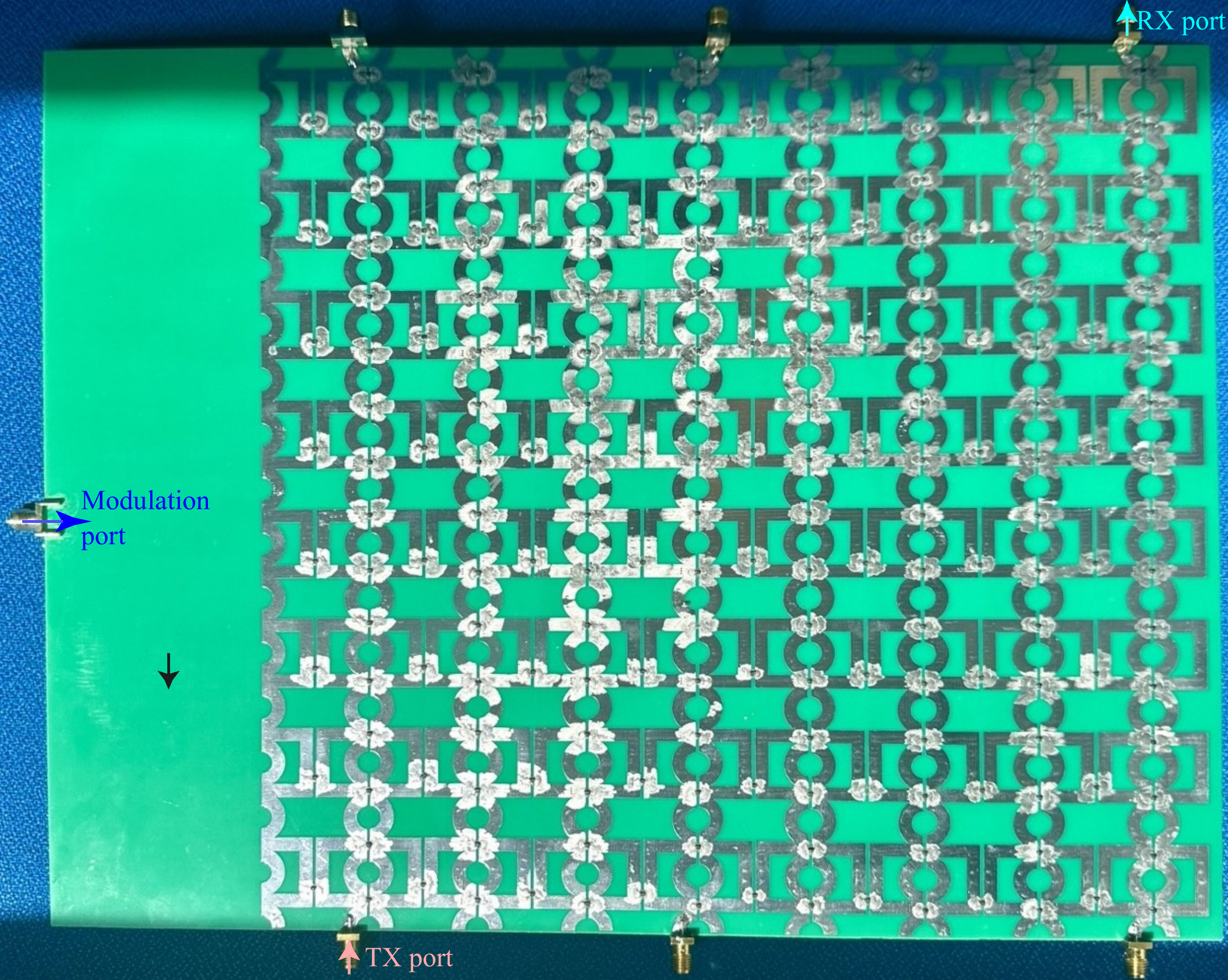}}
		\subfigure[]{\label{Fig:Meas}
			\includegraphics[width=1\columnwidth]{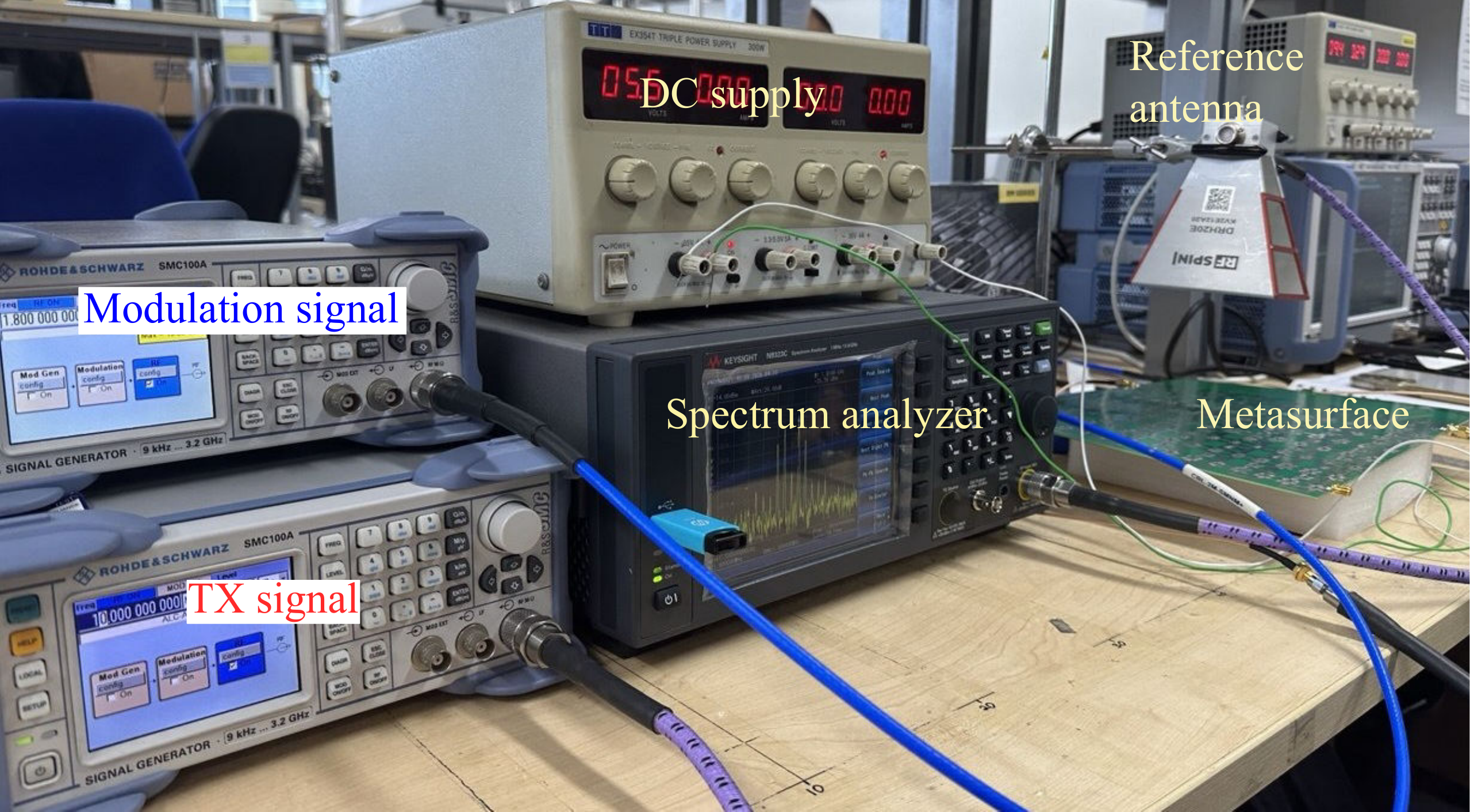}}
		\caption{Photos of the fabricated prototype and measurement set-up. (a)~Top side showing the varactor-loaded unit-cells. (b)~Measurement set-up using a broad-band horn antenna, two signal generators and a spectrum analyser.}
		\label{Fig:Photo}
	\end{center}
\end{figure}

To demonstrate the frequency conversion capability of the proposed space-time-modulated bianisotropic metasurface, we fabricated a prototype, shown in Fig.~\ref{Fig:Phototop}. The measurement setup (Fig.~\ref{Fig:Meas}) consisted of a broadband horn antenna, two signal generators, and a spectrum analyzer. The metasurface was biased with a traveling-wave modulation waveform at $\Omega/(2\pi)=1.8$~GHz, propagating from left to right along the structure. The input signal was set to $\omega_\text{tx}/(2\pi) = 10$~MHz.

\begin{figure}
	\begin{center}
		\subfigure[]{\label{Fig:TX3V}
			\includegraphics[width=0.65\columnwidth]{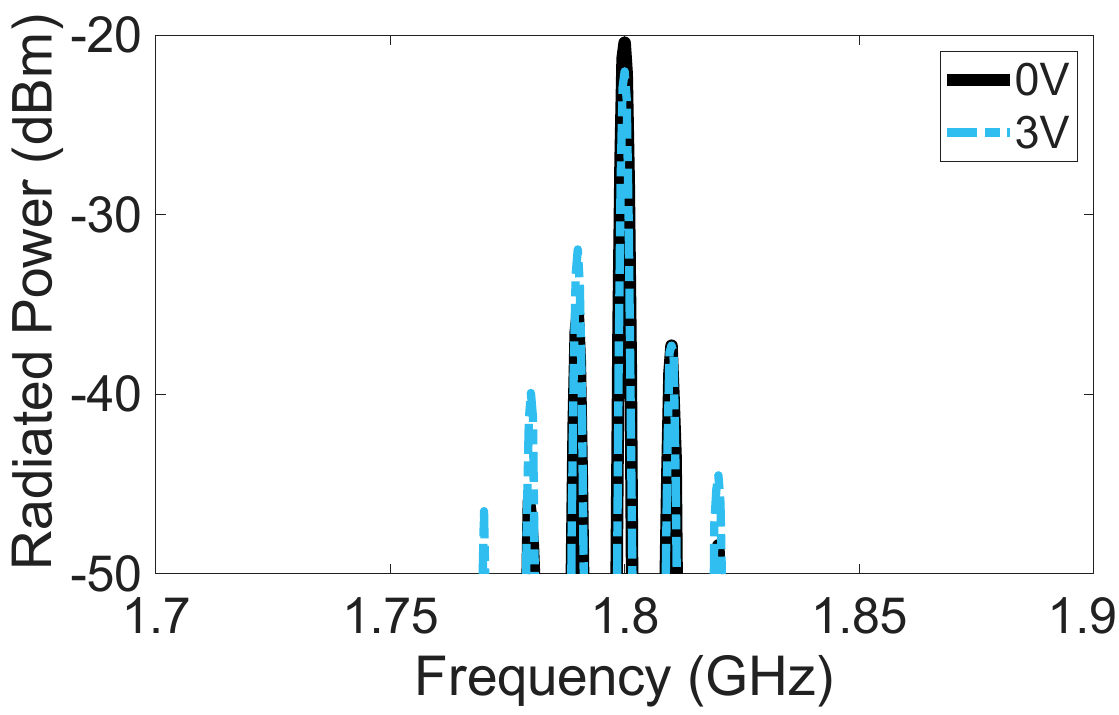}}
		\subfigure[]{\label{Fig:TX4p3V}
			\includegraphics[width=0.65\columnwidth]{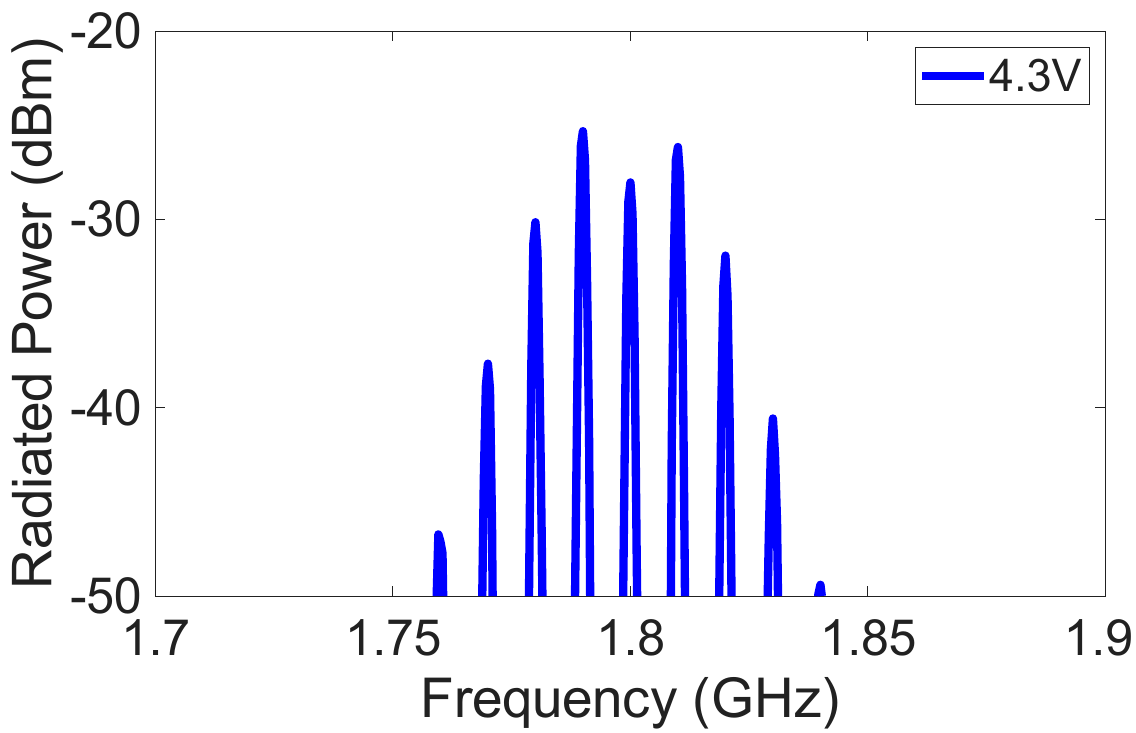}}
		\subfigure[]{\label{Fig:TX6p4V}
			\includegraphics[width=0.65\columnwidth]{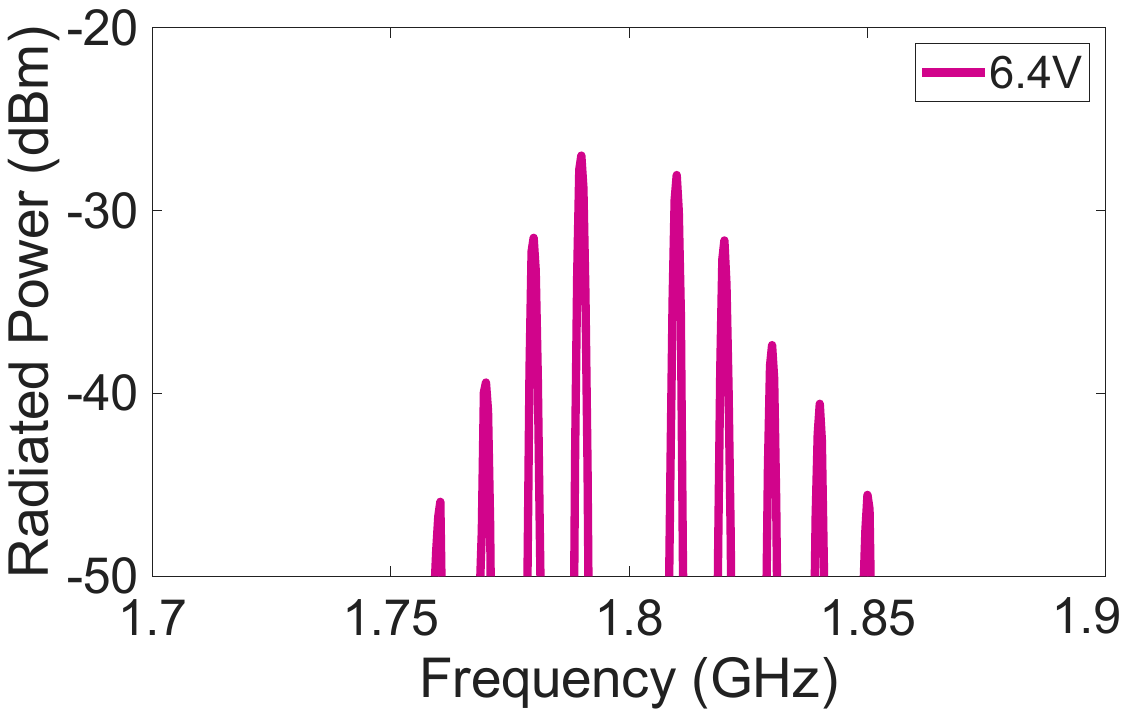}}
		\caption{Bias-controlled polychromatic harmonic generation and radiation. Measured radiated power spectra for DC bias voltages of (a) 0 V and 3 V, (b) 4.3 V, and (c) 6.4 V, with input signal $\omega_\text{tx}/(2\pi)=10$ MHz and modulation frequency $\Omega/(2\pi)=1.8$ GHz. The spectrum broadens from a single dominant tone in (a) to a nine-line comb of sidebands $\omega_{1,m}=\Omega+m\omega_\text{tx}$ in (c).}
		\label{Fig:Radiation}
	\end{center}
\end{figure}

Figure~\ref{Fig:Radiation} shows that the measured spectrum varies strongly with the DC bias applied to the varactors. At $V_\text{DC}=0$ and 3~V [Fig.~\ref{Fig:TX3V}], the 1.8-GHz pump component dominates and the first-order mixing products are more than 25~dB lower. At 4.3~V [Fig.~\ref{Fig:TX4p3V}] and 6.4~V [Fig.~\ref{Fig:TX6p4V}], power is redistributed among components $\omega_{1,m}=\Omega+m\omega_\text{tx}$ with $m=\pm1,\pm2,\pm3,\pm4$. At 6.4~V, the components at 1.79 and 1.81~GHz are prominent and the 1.8-GHz component is substantially reduced. The data establish bias-dependent spectral redistribution, but they do not identify a unique cancellation or ``balanced-mode'' mechanism. Possible contributions include bias-dependent resonance detuning, impedance matching, delivered modulation voltage, and varactor nonlinearity. Because the modulation voltage at the diodes and the bias-dependent $C$--$V$ operating trajectory were not measured, a modulation depth and absolute conversion efficiency are not reported.

\begin{figure}
		\begin{center}
			\subfigure[]{\label{Fig:Receptiona}
				\includegraphics[width=0.48\columnwidth]{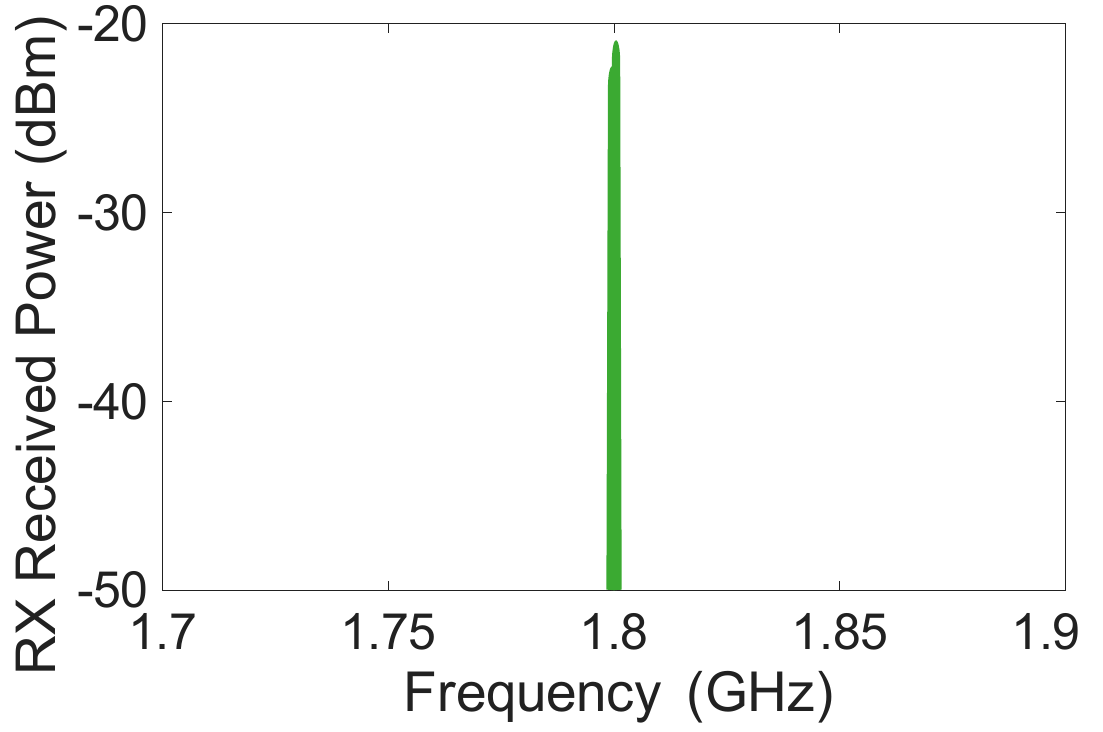}}
			\subfigure[]{\label{Fig:Receptionb}
				\includegraphics[width=0.48\columnwidth]{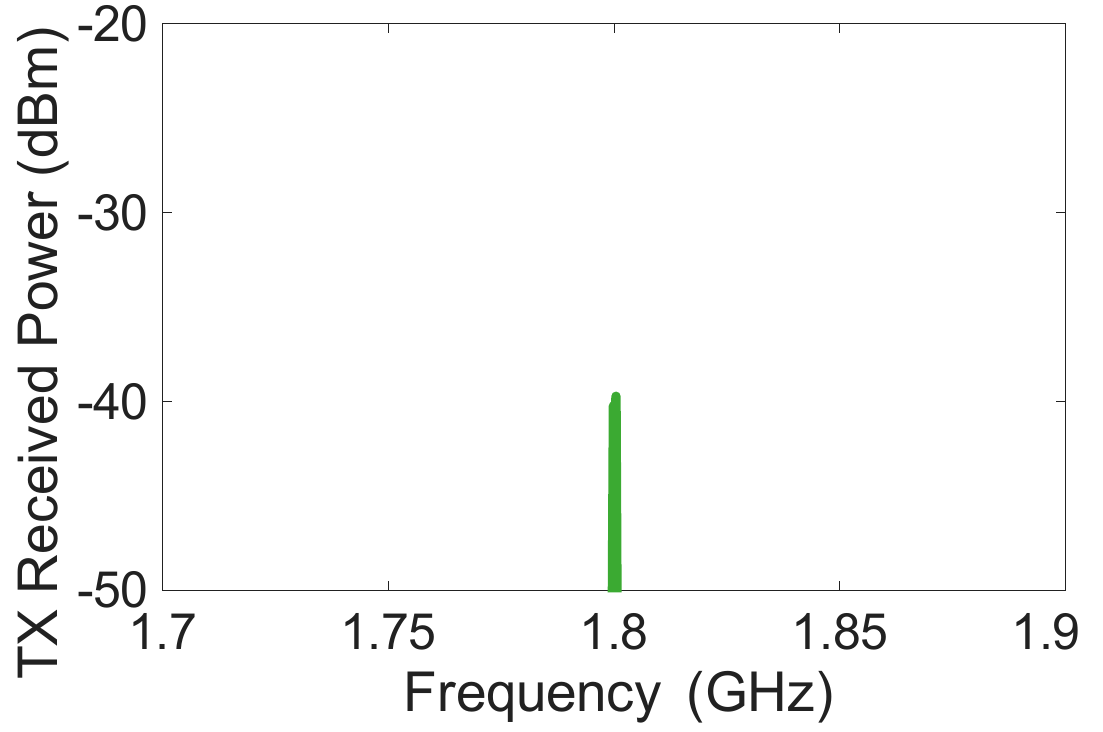}}
			\caption{Directional receive-port routing under free-space illumination. For $\omega_\text{inc}/(2\pi)=1.8$~GHz, the spectra captured at (a) the RX port and (b) the TX port are compared under identical incidence and bias conditions. The 1.8-GHz components peak at $-21$~dBm and $-38$~dBm, respectively, corresponding to an RX-to-TX port-power contrast of approximately 17~dB. This comparison quantifies preferential port coupling; it is not a comparison of a channel with its reversed scattering channel.}
			\label{Fig:Reception}
		\end{center}
	\end{figure}
	
To quantify directional reception, the metasurface is illuminated from free space at $\omega_\text{inc}/(2\pi)=1.8$~GHz and the powers captured at the RX and TX ports are compared under the same incidence and bias conditions. Figure~\ref{Fig:Receptiona} shows a peak of $-21$~dBm at the RX port, whereas Fig.~\ref{Fig:Receptionb} shows $-38$~dBm at the TX port, giving an approximately 17-dB RX-to-TX contrast. The result demonstrates preferential coupling toward the RX port and is consistent with direction-dependent phase matching produced by the traveling modulation. However, the two traces correspond to two different output ports under the same illumination. They are therefore not, by themselves, a complete reciprocity test. Such a test would compare a specified free-space channel to a guided port with the properly normalized reversed process, including the relevant frequency-converted channel, propagation direction, and polarization. We consequently use Fig.~\ref{Fig:Reception} as evidence of directional port routing rather than as standalone proof of electromagnetic nonreciprocity.

\begin{figure}
		\begin{center}
			\subfigure[]{\label{Fig:3Da}
				\includegraphics[width=0.5\columnwidth]{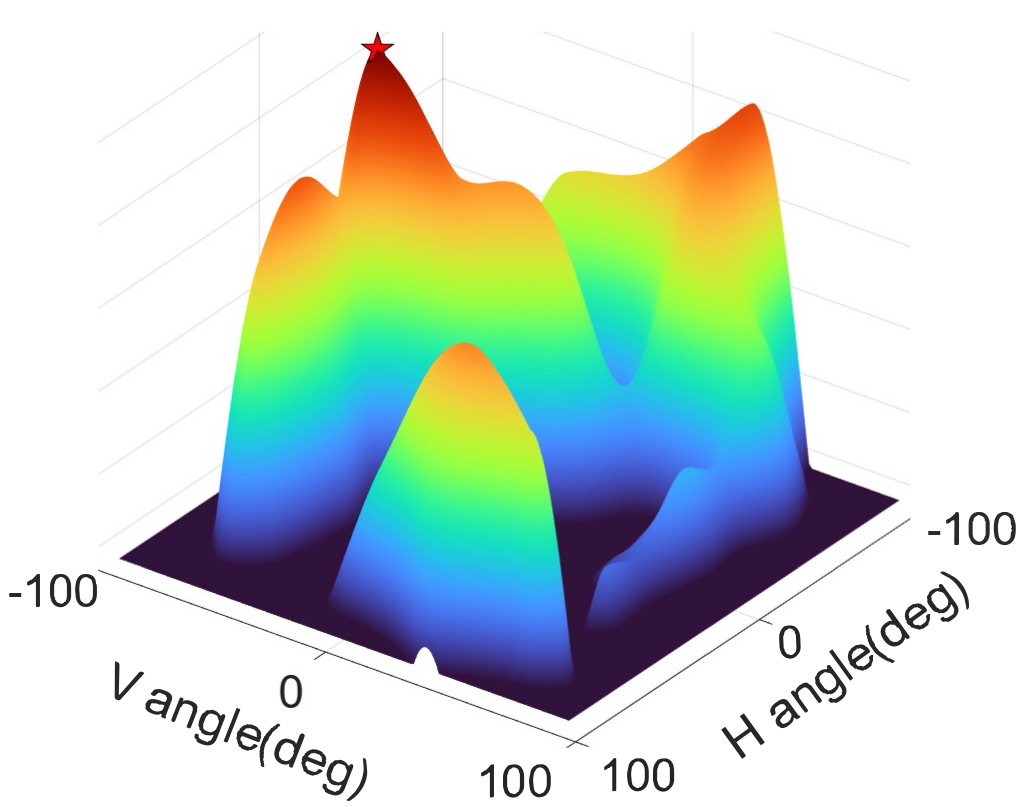}}
			\subfigure[]{\label{Fig:3Db}
				\includegraphics[width=0.45\columnwidth]{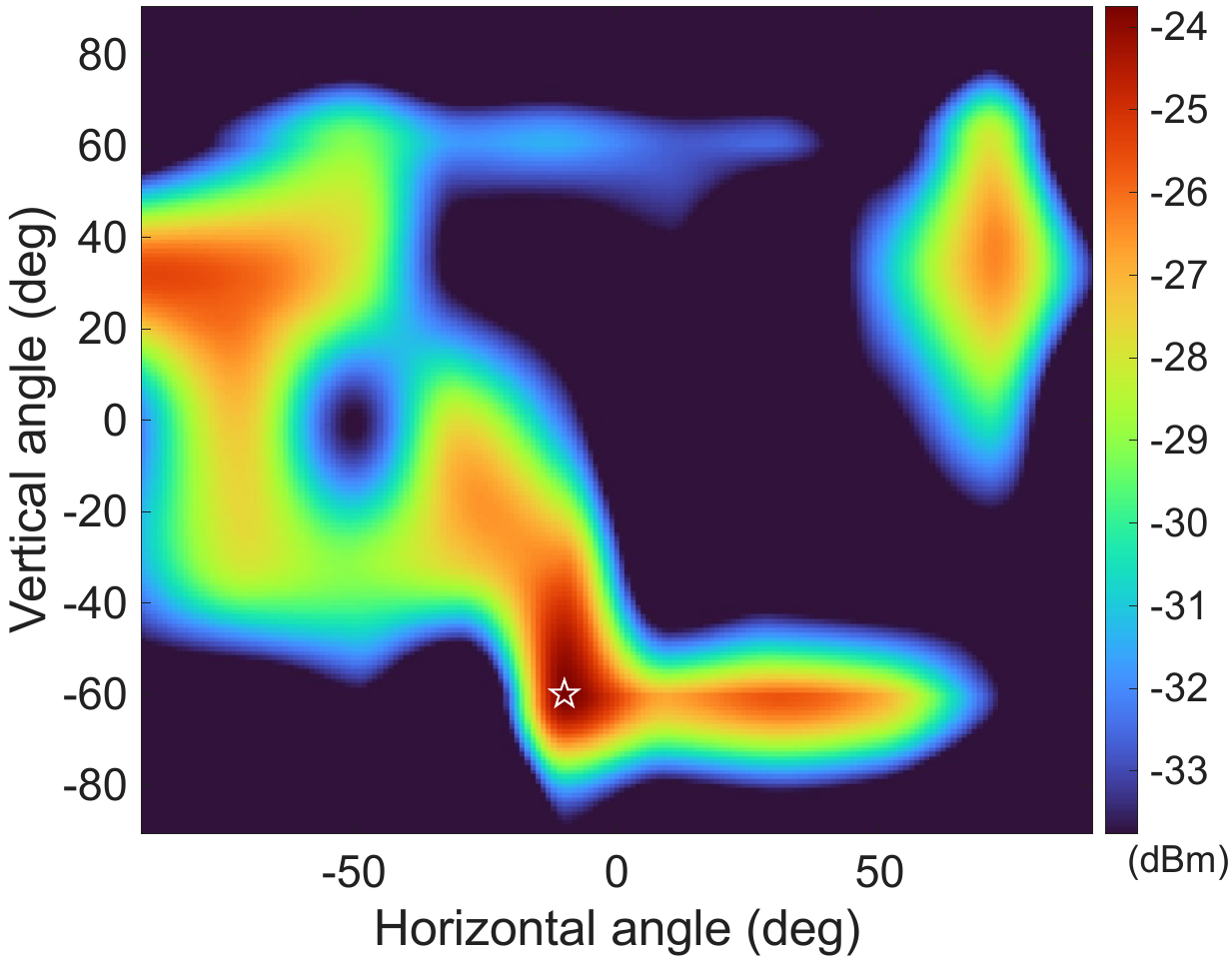}}
			\subfigure[]{\label{Fig:3Dc}
				\includegraphics[width=0.5\columnwidth]{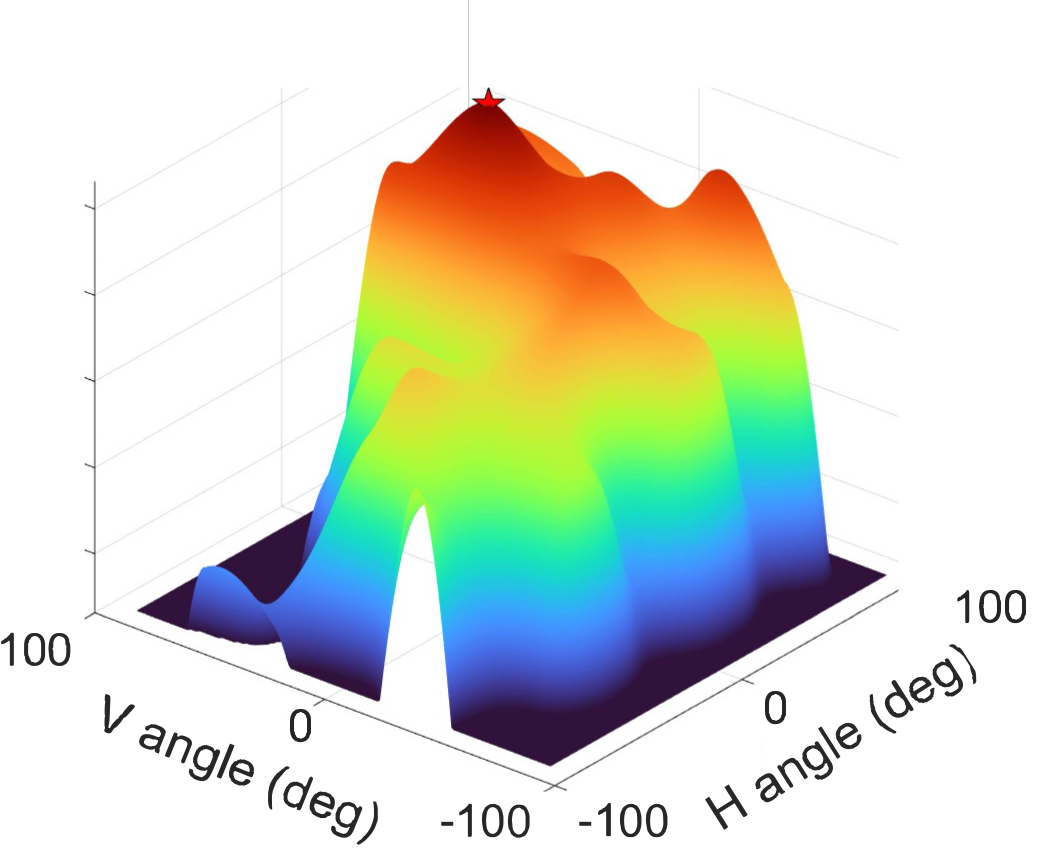}}
			\subfigure[]{\label{Fig:3Dd}
				\includegraphics[width=0.45\columnwidth]{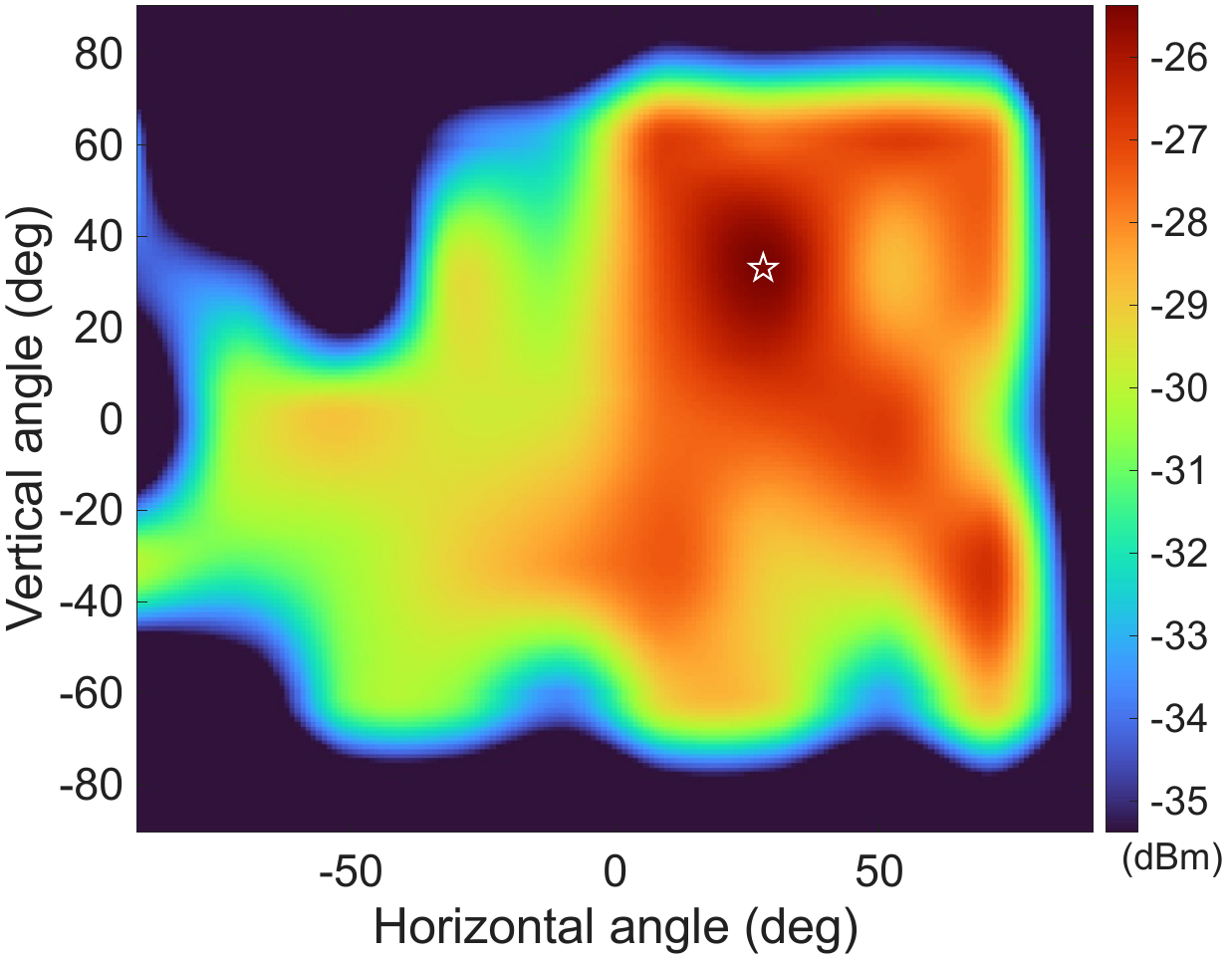}}
			\subfigure[]{\label{Fig:3De}
				\includegraphics[width=0.5\columnwidth]{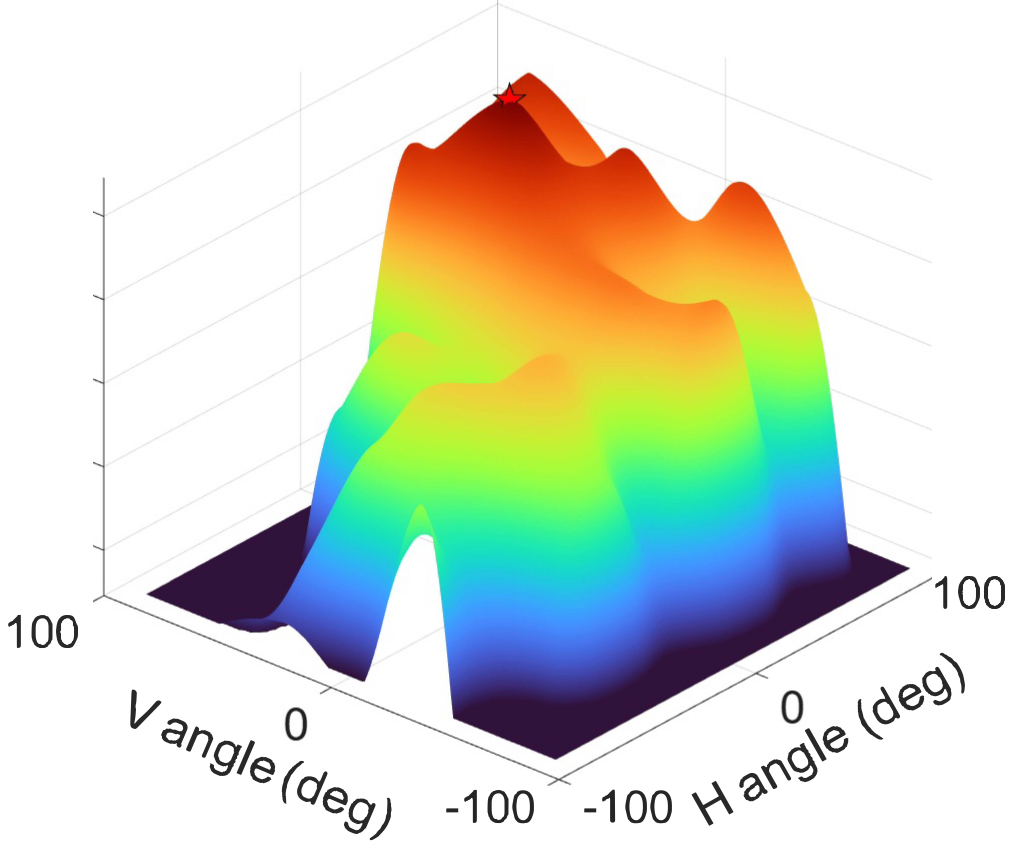}}
			\subfigure[]{\label{Fig:3Df}
				\includegraphics[width=0.45\columnwidth]{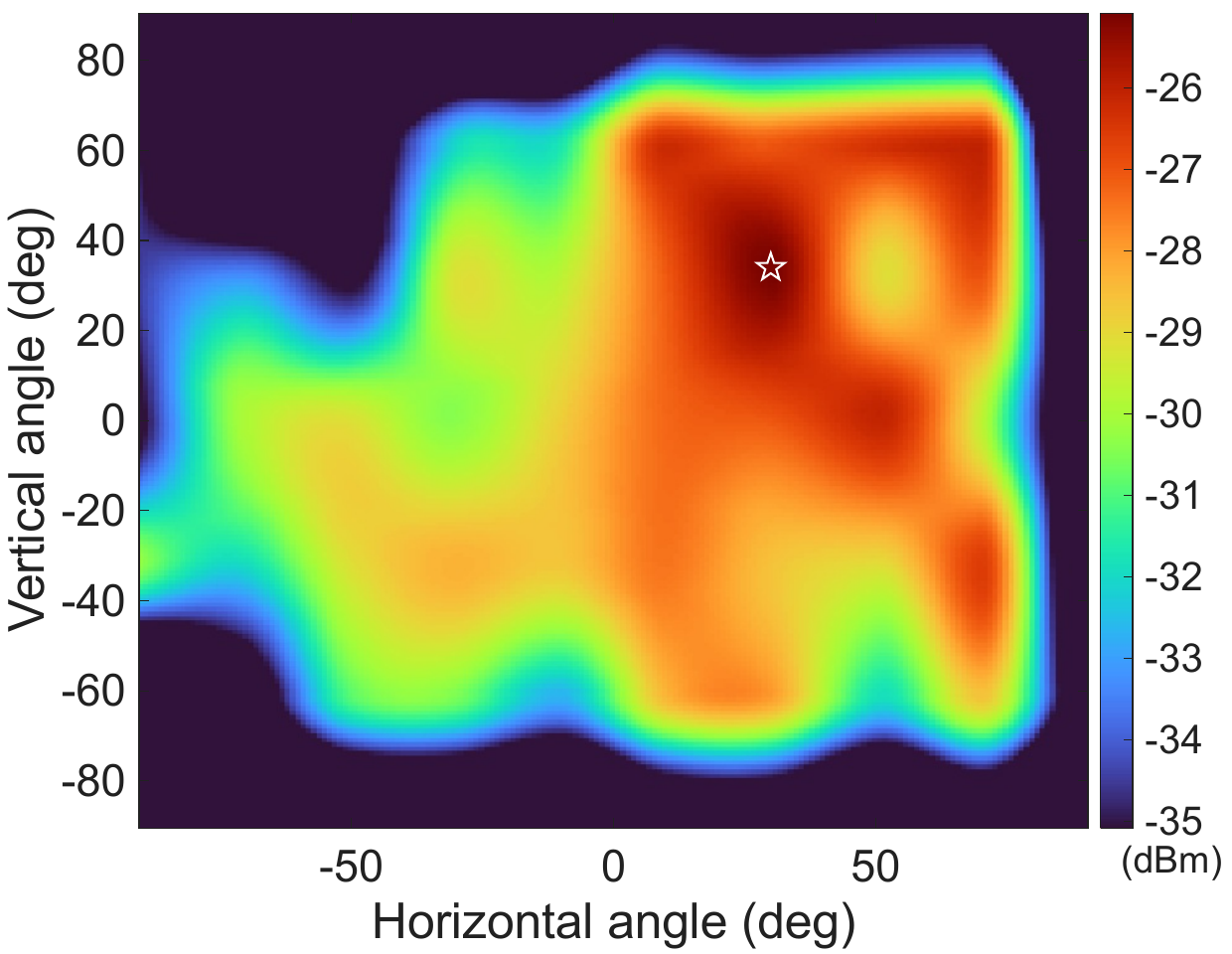}}
			\caption{Measured three-dimensional TX radiation patterns at $\Omega/(2\pi)=1.80$~GHz and at the first sidebands $(\Omega\pm\omega_\text{tx})/(2\pi)=1.81$ and $1.79$~GHz. (a,c,e) Radiation pattern as a function of horizontal (H) and vertical (V) observation angle for 1.80, 1.81, and 1.79~GHz, respectively. Each interpolated pattern is normalized to its own peak and displayed over a 10-dB dynamic range. (b,d,f) Corresponding top-view projections, with the measured peak direction marked by a star. The two first sidebands peak in the same angular region and are not interpreted as independently steered beams.}
			\label{Fig:3D}
		\end{center}
	\end{figure}
	
Figure~\ref{Fig:3D} maps the measured three-dimensional TX radiation patterns of the component at $\Omega/(2\pi)$ and the first sidebands. The sidebands $\omega_{1,+1}=\Omega+\omega_\text{tx}$ and $\omega_{1,-1}=\Omega-\omega_\text{tx}$ [Figs.~\ref{Fig:3Dc}--\ref{Fig:3Df}] both peak near $(V,H)\approx(+40^\circ,+50^\circ)$ and have broad, substantially overlapping main lobes. This behaviour is consistent with Eqs.~\eqref{eq:fast_slow_wave} and~\eqref{eq:sideband_angle_approx}: both products have spatial Floquet order $n=1$, while $\pm\beta_\text{tx}$ and $\pm\omega_\text{tx}$ are small relative to $k_\text{m}$ and $\Omega$, respectively. Hence, $\theta_{1,+1}\approx\theta_{1,-1}$ is expected, and the two sidebands occupy a common angular region rather than forming independently steerable beams.

The measured component at $\Omega/(2\pi)$ [Figs.~\ref{Fig:3Da} and~\ref{Fig:3Db}] peaks in a different region, near $(V,H)\approx(-60^\circ,0^\circ)$. The reduced momentum relation alone does not explain this large difference if the $\Omega/(2\pi)$ component is treated simply as the $(n,m)=(1,0)$ member of the same Floquet family. The measured $\Omega/(2\pi)$ pattern may also contain direct pump radiation and a different aperture-current distribution from those producing the converted sidebands. Because these contributions were not independently separated, Fig.~\ref{Fig:3D} is used to report the measured directional radiation of the generated spectrum and the theoretically expected near-coincidence of the two first-sideband angles, not to claim independent beam steering or a quantitative prediction of the $\Omega/(2\pi)$ peak direction.

All modulated measurements were performed near 1.8~GHz with a broadband horn antenna and a single bias network feeding $\text{D}_1$, $\text{D}_2$, and $\text{D}_3$. The horn provides both electric and magnetic field components as part of the incident electromagnetic wave. The simulations in Figs.~\ref{Fig:Esim} and~\ref{Fig:Hsim} show that the complete unit cell supports coexisting electric and ring-associated magnetic responses at this frequency. Nevertheless, no experiment disables either varactor set, and the modulated susceptibility components are not retrieved. The measured spectra therefore cannot be assigned quantitatively to the electric, magnetic, or magnetoelectric channels individually. A stronger separation would require electric-only, ring-only, and combined-modulation controls, together with static and modulated scattering-parameter measurements and susceptibility retrieval. These limitations do not affect the directly measured polychromatic spectra, radiation patterns, or RX-to-TX port-power contrast, but they bound the associated physical claims.

\section{Conclusion}
\label{sec:conclusion}
We have demonstrated a bianisotropic space-time-modulated metasurface composed of varactor-loaded omega-shaped unit cells. Full-wave simulations reveal coexisting electric and magnetic responses at the operating frequency, including strong transverse electric fields and pronounced magnetic-field localization within the circular rings. Traveling-wave modulation of varactors embedded in these responsive regions makes the composite bianisotropic response periodic in both space and time. Experimentally, the modulation generates a bias-reconfigurable polychromatic spectrum containing multiple frequency components with harmonic-dependent directional radiation. The first sidebands radiate into approximately the same angular region, consistent with their common spatial Floquet order. Under free-space illumination, the metasurface also exhibits preferential coupling toward the RX port, demonstrating modulation-controlled directional routing. These results establish bianisotropic space-time metasurfaces as a platform in which electric, magnetic, spectral, and spatial degrees of freedom can be jointly manipulated within a single dynamically modulated aperture. The demonstrated combination of polychromatic radiation and directional reception opens opportunities for reconfigurable wireless front ends, frequency-multiplexed apertures, and multifunctional electromagnetic surfaces.
	
\bibliography{Taravati_Reference}
	
\end{document}